\pdfoutput=1
\documentclass[11pt]{article}
\usepackage[margin=2.6cm]{geometry}
\usepackage[english]{babel} %tildes etc
\usepackage[utf8]{inputenc}
\usepackage{indentfirst}
\usepackage{amsmath}
\usepackage{amssymb}
\usepackage{url}
\usepackage{graphicx}
\usepackage{braket}
\usepackage{slashed}
\usepackage{cancel}
\usepackage[font=footnotesize,labelfont=bf]{caption}
\usepackage[usenames, dvipsnames]{xcolor}
\usepackage{booktabs}
\usepackage[colorlinks=true, urlcolor=MidnightBlue, linkcolor=MidnightBlue, citecolor=Mahogany, pdfborder={0 0 0}]{hyperref}
\usepackage[sort&compress,numbers]{natbib}
\usepackage{doi}%<----------
\usepackage[capitalise]{cleveref}
\let\emph\relax % there's no \RedeclareTextFontCommand
\DeclareTextFontCommand{\emph}{\color{MidnightBlue}\em}
\DeclareTextFontCommand{\comm}{\color{BurntOrange}\em}
\newcommand{\CP}{$C\!P$}
\newcommand{\M}[2]{\mathcal{M}^{\textrm{#1}}_{#2}}
\newcommand{\MM}[2]{\mathcal{M}^{\textrm{#1}}_{\textrm{#2}}}
\newcommand{\ExpPi}{Samios:1962zza,Abouzaid:2008cd}
\newcommand{\ExpEta}{KLOE2:2011aa}%remove upper limits results
\newcommand{\ExpKe}{Barr:1994wc,Akagi:1994bb,Vagins:1993ws,Gu:1994nb,AlaviHarati:2001ab,Lai:2000ic}%remove upper limits results and superseeded ones
\newcommand{\ExpKemu}{Gu:1996tu,AlaviHarati:2002eh}%remove upper limits results and superseeded ones
\title{\Large\bfseries{The radiative corrections to double-Dalitz decays revisited}}
\author{\normalsize{
        Karol Kampf{\color{Mahogany}\thanks{kampf@ipnp.troja.mff.cuni.cz}}, 
        Ji\v{r}i Novotn\'y{\color{Mahogany}\thanks{novotny@ipnp.troja.mff.cuni.cz}}, 
        Pablo Sanchez-Puertas{\color{Mahogany}\thanks{sanchezp@ipnp.troja.mff.cuni.cz}}
        } \vspace{0.2cm}  \\ 
        {\small{Faculty of Mathematics and Physics, Institute of Particle and Nuclear Physics,}} \\
        {\small{Charles University in Prague, V Holešovičkách 2, 18000 Praha 8, Czech Republic}}
}
\date{}
\begin{document}\renewcommand{\abstractname}{\vspace{-\baselineskip}} \maketitle
\begin{abstract} 

In this study, we revisit and complete the full next-to-leading order corrections to pseudoscalar double-Dalitz decays within the soft photon approximation. Comparing to the previous study, we find small differences, which are nevertheless relevant for extracting information about the pseudoscalar transition form factors. Concerning the latter, these processes could offer the opportunity to test them---for the first time---in their double-virtual regime.

\end{abstract}
%%%%%%% IF USING TWO-COLUMNS: GENERATE TITLE AND ABSTRACT  %%%%%%
%\twocolumn[
%  \begin{@twocolumnfalse}
%    \maketitle
%    \begin{abstract}
%      This is the abstract. This is the abstract. This is the abstract. This is the abstract. This is the abstract. This is the abstract. 
%      This is the abstract. This is the abstract. This is the abstract. This is the abstract.\\
%    \end{abstract}
%  \end{@twocolumnfalse}
%]
%%%%%%%%%%%%%%%%%%%%%%%%%%%%%%%%%%%%%%%%%%%%%%%%%%%%%%%%%%%%%%%%%

%%%%%%%%%%%%%%%%%%%%%%%%%%%%%%%%%%%%%%%%%%%%%%%%%%%%%%%%%%%%%%%%%
%%%%%%%%%%%%%%%%%%%%%%%%%%%%%%%%%%%%%%%%%%%%%%%%%%%%%%%%%%%%%%%%%
%%%%%%%%%%%%%%%%%%%%%                    %%%%%%%%%%%%%%%%%%%%%%%%
%%%%%%%%%%%%%%%%%%%%%    INTRODUCTION    %%%%%%%%%%%%%%%%%%%%%%%%
%%%%%%%%%%%%%%%%%%%%%                    %%%%%%%%%%%%%%%%%%%%%%%%
%%%%%%%%%%%%%%%%%%%%%%%%%%%%%%%%%%%%%%%%%%%%%%%%%%%%%%%%%%%%%%%%%
%%%%%%%%%%%%%%%%%%%%%%%%%%%%%%%%%%%%%%%%%%%%%%%%%%%%%%%%%%%%%%%%%
\section{\large{Introduction}}

Double-Dalitz decays of pseudoscalar mesons ($P\to \bar\ell\ell \bar\ell'\ell'$) have attracted attention over the years, both theoretically~\cite{Kroll:1955zu,Miyazaki:1974qi,Uy:1990hu,Bijnens:1999jp,Barker:2002ib,Lih:2009np,Petri:2010ea,Terschlusen:2013iqa,D'Ambrosio:2013cqa,Escribano:2015vjz,Weil:2017knt} and experimentally~\cite{\ExpPi,\ExpEta,\ExpKe,\ExpKemu}, for different reasons. On the one hand, they contain important---direct---information about the pseudoscalar meson structure, which is encoded in their double-virtual transition form factors (TFFs). Interesting enough, double-virtual effects have never been measured, and are relevant for predicting the hadronic light-by-light contribution to the anomalous magnetic moment of the muon~\cite{Jegerlehner:2009ry,Masjuan:2017tvw}. On the other hand, the angular distribution associated to the lepton planes ($\phi\equiv\phi_{\bar\ell\ell, \bar\ell'\ell'}$) is a \CP{}-sensitive observable and was indeed the first experimental evidence for the parity of the $\pi^0$~\cite{Plano:1959zz,Samios:1962zza}. Since no significant amount of \CP{} violation is expected in these processes within the standard model, any signal of this would be very interesting\footnote{With the possible exception of $K_L$ decays, see Ref.~\cite{AlaviHarati:1999ff}.}. However, before extracting any information from these decays, a careful analysis of the next-to-leading-order (NLO) radiative corrections (RCs) is required as we shall see. 
A partial analysis of the NLO RC was performed in Ref.~\cite{Barker:2002ib}, finding sizeable corrections. In this study, we review the RC evaluated in Ref.~\cite{Barker:2002ib} and include their missing diagrams in order to obtain the full NLO corrections. 

The paper is structured as follows: the leading-order (LO) results and definitions are presented in \cref{sec:LO}, whereas the NLO corrections are introduced in \cref{sec:NLO}---which includes the new corrections as well as analytical and numerical comparison to previous results in Ref.~\cite{Barker:2002ib}. 
Finally, in \cref{sec:TFFeff}, we discuss briefly about experimental prospects regarding TFFs.

%%%%%%%%%%%%%%%%%%%%%%%%%%%%%%%%%%%%%%%%%%%%%%%%%%%%%%%%%%%%%%%%%
%%%%%%%%%%%%%%%%%%%%%%%%%%%%%%%%%%%%%%%%%%%%%%%%%%%%%%%%%%%%%%%%%
%%%%%%%%%%%%%%%%%%%%%                    %%%%%%%%%%%%%%%%%%%%%%%%
%%%%%%%%%%%%%%%%%%%%%     LO RESULTS     %%%%%%%%%%%%%%%%%%%%%%%%
%%%%%%%%%%%%%%%%%%%%%                    %%%%%%%%%%%%%%%%%%%%%%%%
%%%%%%%%%%%%%%%%%%%%%%%%%%%%%%%%%%%%%%%%%%%%%%%%%%%%%%%%%%%%%%%%%
%%%%%%%%%%%%%%%%%%%%%%%%%%%%%%%%%%%%%%%%%%%%%%%%%%%%%%%%%%%%%%%%%
\section{\large{LO Results}}\label{sec:LO}

\begin{figure}[t]
\centering
  \includegraphics[width=0.6\textwidth]{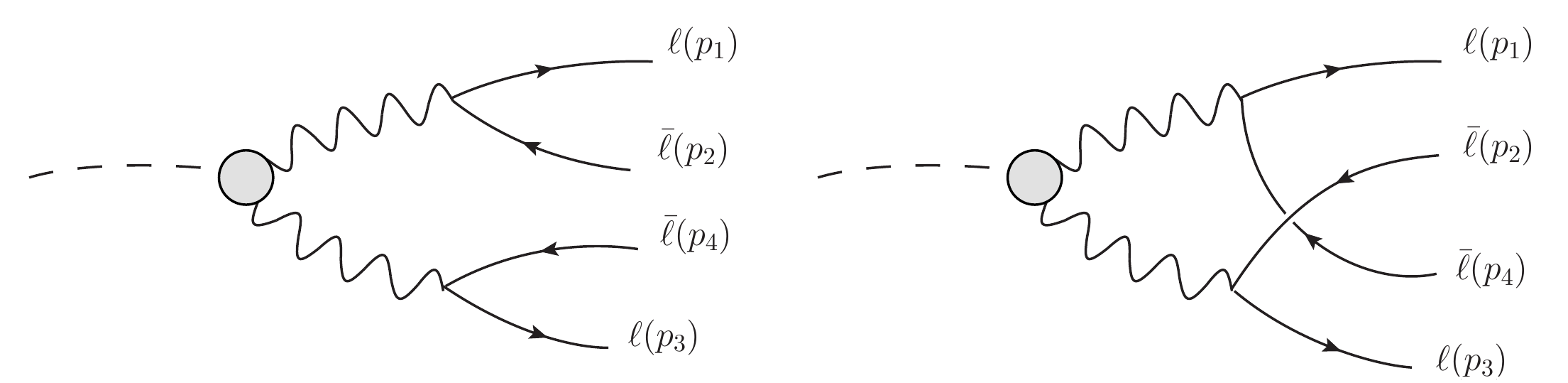}
\caption{The direct (left) and exchange (right) diagrams contributing to the process (the latter one appears for identical fermions in the final state).\label{fig:LO}}
\end{figure}

The LO result is given by the tree-level processes depicted in \cref{fig:LO} (left) [for identical leptons an additional---exchange---diagram appears, see \cref{fig:LO} (right)], which amplitude is related to the anomaly.\footnote{We use $\epsilon^{0123} =+1$; see \cref{sec:kin,sec:cpv} regarding conventions, (effective) Lagrangians, and matrix elements.} Particularly, for the direct and exchange contribution we obtain
  \begin{align}
    i\mathcal{M}^{\textrm{LO}}_D &= -ie^4\frac{F_{P\gamma\gamma}(s_{12},s_{34})}{s_{12} s_{34}}
                                     \epsilon_{\mu\nu\rho\sigma}p_{12}^{\mu}p_{34}^{\rho} 
                                     \left[ \bar{u}(p_1)\gamma^{\nu}v(p_2)\right]  \left[\bar{u}(p_3)\gamma^{\sigma}v(p_4) \right], \\
    i\mathcal{M}^{\textrm{LO}}_E &= + ie^4\frac{F_{P\gamma\gamma}(s_{14},s_{32})}{s_{14} s_{32}} 
                                     \epsilon_{\mu\nu\rho\sigma}p_{14}^{\mu}p_{32}^{\rho} 
                                     \left[ \bar{u}(p_1)\gamma^{\nu}v(p_4)\right]  \left[\bar{u}(p_3)\gamma^{\sigma}v(p_2) \right],
  \end{align}
respectively, where $F_{P\gamma\gamma}(q_1^2,q_2^2)$ is the pseudoscalar TFF and encodes the meson structure. Note, in particular, the relative sign for the exchange contributions, which is generic and arises from Fermi statistics. The amplitude squared can be expressed then as a combination of direct ($\vert\M{LO}{D}\vert^2$), exchange ($\vert\M{LO}{E}\vert^2$) and interference ($2\operatorname{Re}\M{LO}{D}\M{LO$*$}{E}$) terms. Employing the Cabibbo Maskimowicz description~\cite{Cabibbo:1965zzb} for the four-body final state (see \cref{sec:kin}), these read
  \begin{multline}\label{eq:ampdir}
    |\M{LO}{D}|^2 = \frac{e^8 |F_{P\gamma\gamma}(s_{12},s_{34})|^2}{x_{12}x_{34}}\lambda^2 
    \Big( 2 - \lambda_{12}^2 + y_{12}^2 -\lambda_{34}^2 + y_{34}^2 +(\lambda_{12}^2 - y_{12}^2)(\lambda_{34}^2 - y_{34}^2)\sin^2\phi\Big),
  \end{multline}
  \begin{multline}\label{eq:ampexc}
   |\M{LO}{E}|^2 = \frac{e^8 |F_{P\gamma\gamma}(s_{14},s_{32})|^2}{x_{14}x_{32}} \Bigg[
    \lambda^2_{ex}\Big( 2 - (\lambda_{14}^2 + \lambda_{32}^2) \Big) + 2(x_{12}-x_{34})^2 \\ 
    + \lambda^2\Big( \frac{1}{2}(y_{12}+y_{34})^2  + (\lambda_{12}^2 - y_{12}^2)(\lambda_{34}^2 - y_{34}^2)\sin^2\phi \  \frac{x_{12}x_{34}}{x_{14}x_{32}} \Big)  \Bigg],
  \end{multline}
  \begin{multline}
    2\operatorname{Re}\M{LO}{D}\M{LO*}{E} = 
    \frac{e^8 \operatorname{Re}F_{P\gamma\gamma}(s_{12},s_{34}) F^*_{P\gamma\gamma}(s_{14},s_{32})}{8x_{12}x_{34}x_{14}x_{32}} \lambda^2 
    \Big(
    8\eta^4 - w^2(1+y_{12}y_{34})(2-y_{12}^2-y_{34}^2) \\
    + 4\eta^2(x_{12}y_{12}+x_{34}y_{34})(y_{12}+y_{34})  + (8\eta^2-z(y_{12}+y_{34})^2)\Xi +2\Xi^2
    \Big),
  \end{multline}
which are in good agreement with Ref.~\cite{Barker:2002ib}. Exchange contributions, such as \cref{eq:ampexc}, can be obtained, in general, from the direct ones by shifting to the exchange variables, a procedure which is much more efficient and is outlined in \cref{sec:kin}. 

Finally in this section, we obtain the double-Dalitz branching ratios in terms of the two-photons decay ($\Gamma_{4\ell}/\Gamma_{2\gamma}$) for different pseudoscalars and lepton species considering both, the case of a constant TFF, and a simple---but precise low-energy---TFF description in terms of Pad\'e approximants described in \cref{sec:TFF}.  The decay widths are given, in general, by\footnote{See \cref{sec:kin} for the phase-space boundaries and $d\Phi_4$ definitions.}
  \begin{equation}
    \Gamma_{4\ell} = \frac{1}{2M}\int d\Phi_4 \vert\M{}{D}\vert^2 + \left(\vert\M{}{E}\vert^2 + 2\operatorname{Re}\M{}{D}\M{*}{E} \right), 
    \qquad \Gamma_{2\gamma} = \frac{\pi\alpha^2M^3}{4}|F_{P\gamma\gamma}(0,0)|^2.
  \end{equation}
Note in particular that direct and exchange terms contribute the same to the total decay width, and it is therefore sufficient to calculate the direct one. Furthermore, we introduce a change of variables that improves the numerical integration convergence and proves valuable when calculating the NLO contributions: 
  \begin{equation}\label{eq:cvarslog}
    s_{12(34)} \rightarrow 4m_{a(b)}^2 \operatorname{exp}\frac{\bar{s}_{12(34)}}{4m_{a(b)}^2}, \quad
    ds_{12}ds_{34} \to d\bar{s}_{12}d\bar{s}_{34} \frac{s_{12}s_{34}}{16m_a^2m_b^2} \ . 
  \end{equation}
This cancels out the photon propagators in $\M{LO}{D}$, resulting in a flatter---non peaked---integrand.\footnote{We expect this change of variables to be valuable for Monte Carlo (MC) generators that would require us to evaluate many events in the hit-or-miss procedure otherwise.} We quote our LO results in  \cref{tab:LO} with the only exception of the $\eta'$, which we postpone for a future work. The reason for this is the presence of resonant structures for the electronic modes that require certain care when describing the TFF---especially if dealing with NLO corrections (see Ref.~\cite{Husek:2017vmo} in this respect for the Dalitz decay case). 
  \begin{table}[tbp]\centering\small
  \begin{tabular}{cccccccc}\toprule
           & $\pi^0\to 4e$ & $K_L\to 4e$ & $K_L\to 2e2\mu$ & $K_L\to 4\mu$ & $\eta\to 4e$ & $\eta\to 2e2\mu$ & $\eta\to 4\mu$  \\ \midrule
  D+E      &   $\phantom{-}3.4558(3)$ & $\phantom{-}6.2582(6)$ & $2.8589(3)$ & $\phantom{-}0.9886(1)$ & $\phantom{-}6.4972(6)$ & $3.9961(4)$ & $\phantom{-}6.5622(7)$ \\ 
  Int      &   $-0.0362(3)$           & $-0.0363(4)$           & ---         & $-0.0511(1)$           & $-0.0362(4)$           & ---         & $-0.4883(7)$ \\ 
  Total    &   $\phantom{-}3.4196(4)$ & $\phantom{-}6.2219(7)$ & $2.8589(3)$ & $\phantom{-}0.9375(1)$ & $\phantom{-}6.4610(7)$ & $3.9961(4)$ & $\phantom{-}6.0739(10)$ \\ \midrule
    %%%%%%%%%%%%%%%%%%%%%%%%%%%%%%%%%%%%%%%%%%%%%%%%%%%%%%%%%%%%%%
  FF$_{\textrm{D+E}}$   &   $\phantom{-}3.4692(3)$ & $\phantom{-}6.7457(7)$ & $4.8435(5)$ & $\phantom{-}1.8417(2)$ & $\phantom{-}6.9068(7)$ & $5.9259(6)$ & $\phantom{-}10.658(1)$ \\ 
  FF$_{\textrm{Int}}$   &   $-0.0369(4)$           & $-0.0578(6)$           & ---         & $-0.0972(1)$           & $-0.0537(5)$           & ---         & $-0.818(1)$ \\ 
  FF$_{\textrm{Total}}$ &   $\phantom{-}3.4323(5)$ & $\phantom{-}6.6879(9)$ & $4.8435(5)$ & $\phantom{-}1.7445(2)$ & $\phantom{-}6.8531(9)$ & $5.9259(6)$ & $\phantom{-}9.841(1)$ \\ \bottomrule
  \end{tabular}\caption{$\Gamma_{4\ell}/\Gamma_{2\gamma}$ in units of $10^{-5}$, $10^{-6}$ and $10^{-9}$ for the $4e$, $2e2\mu$ and $4\mu$ modes. The second and third row stand for the sum of direct and exchange (D+E) and interference (Int) terms, respectively; the third row (Total) is the sum of both. The following rows correspond to the analogous result for the TFFs described in \cref{sec:TFF}.\label{tab:LO}}
  \end{table}
The integrals have been performed numerically using the \texttt{CUBA} library~\cite{Hahn:2004fe} and statistical errors are associated to the MC procedure alone\footnote{Furthermore, for the LO calculation, the result was checked with the \texttt{NIntegrate} routine in \texttt{Mathematica}.} and are in good agreement with Ref.~\cite{Barker:2002ib}. Having introduced the main concepts, we move on to the NLO results.

%%%%%%%%%%%%%%%%%%%%%%%%%%%%%%%%%%%%%%%%%%%%%%%%%%%%%%%%%%%%%%%%%
%%%%%%%%%%%%%%%%%%%%%%%%%%%%%%%%%%%%%%%%%%%%%%%%%%%%%%%%%%%%%%%%%
%%%%%%%%%%%%%%%%%%%%%                     %%%%%%%%%%%%%%%%%%%%%%%
%%%%%%%%%%%%%%%%%%%%%     NLO RESULTS     %%%%%%%%%%%%%%%%%%%%%%%
%%%%%%%%%%%%%%%%%%%%%                     %%%%%%%%%%%%%%%%%%%%%%%
%%%%%%%%%%%%%%%%%%%%%%%%%%%%%%%%%%%%%%%%%%%%%%%%%%%%%%%%%%%%%%%%%
%%%%%%%%%%%%%%%%%%%%%%%%%%%%%%%%%%%%%%%%%%%%%%%%%%%%%%%%%%%%%%%%%
\section{\large{Radiative Corrections}}\label{sec:NLO}

At the NLO in $\alpha$, additional amplitudes ($\M{NLO}{}$) appear, resulting in further contributions of the kind
  \begin{align}
     |\mathcal{M}|^2 & = \textrm{LO} + 2\operatorname{Re}\M{NLO}{D}\M{LO*}{D} 
     + 2\operatorname{Re}\left( \M{NLO}{E}\M{LO*}{E} + \M{NLO}{D}\M{LO*}{E} + \M{NLO}{E}\M{LO*}{D}  \right) + \mathcal{O}(\alpha^6) \nonumber \\
     & \equiv \textrm{LO} + \operatorname{Dir}^{\textrm{NLO}} + 
    (\operatorname{Exc}^{\textrm{NLO}} + \operatorname{Int}_D^{\textrm{NLO}} + \operatorname{Int}_E^{\textrm{NLO}}) + \mathcal{O}(\alpha^6), \label{eq:nlogen}
  \end{align}
with obvious identifications.\footnote{In the following, we comment on Dir$^{\textrm{NLO}}$ and Int$^{\textrm{NLO}}_D$ alone---the remaining contributions can be trivially obtained upon the use of the exchange variables defined in \cref{sec:kin}.} The different contributions correspond to, on the one hand, the (TFF-independent) vacuum polarization (\cref{sec:VP}) and vertex functions (\cref{sec:vertex}) and, on the other hand, the additional (TFF-dependent) three-, four-, and five-point loop amplitudes (\cref{sec:3p,sec:4p,sec:5p}). Among the latter, only the five-point was considered in Ref.~\cite{Barker:2002ib}. Therefore, our work completes the---so far missing---full NLO corrections. 
Besides, some terms contain infrared (IR) divergencies that require the inclusion of real photon emission terms; these are the bremsstrahlung (BS) contributions that we account for in the soft-photon approximation in analogy to Ref.~\cite{Barker:2002ib} (\cref{sec:BS}).\footnote{Ref.~\cite{Barker:2002ib} includes also the radiative $P\to \bar{\ell}\ell\bar{\ell}'\ell'\gamma$ decay---besides the soft-photon approximation---for photon energies above certain threshold. In this study, we focus in the purely virtual corrections, for which only the soft-photon contribution is required.} When giving our numerical results, we opt for combining the NLO results with the corresponding BS contribution to obtain a finite IR result. 
In the following, we recapitulate the results from each contribution, commenting on the differences we find with respect to Ref.~\cite{Barker:2002ib}. The numerical results and comparison are relegated to \cref{sec:nlofull}.

%%%%%%%%%%%%%%%%%%%%%%%%%%%%%%%%%%%%%%%%%%%%%%%%%%%%%%%%%%%%%%%%%
%%%%%%%%%%%%%%%     NLO: BREMSSTRAHLUNG     %%%%%%%%%%%%%%%%%%%%%
%%%%%%%%%%%%%%%%%%%%%%%%%%%%%%%%%%%%%%%%%%%%%%%%%%%%%%%%%%%%%%%%%
\subsection{\large{Soft-photon emission}}\label{sec:BS}

\begin{figure}[t]\centering
    \includegraphics[width=0.7\textwidth]{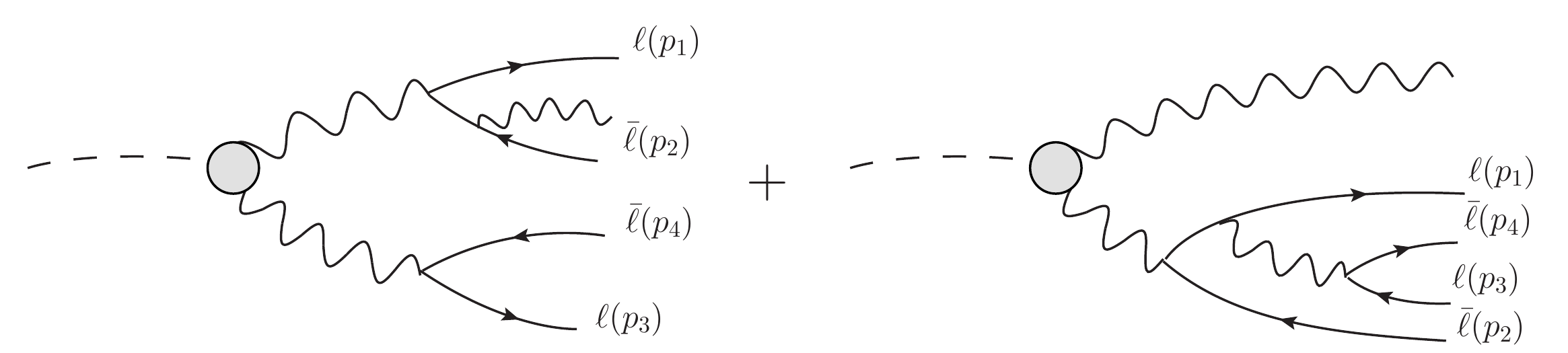}
    \caption{Photon emission graphs (alternative insertions are implicit). In the soft-photon approximation, only the left diagram contributes 	(for identical leptons, exchange diagrams are implied).\label{fig:BS}}
  \end{figure}

The photon emission graphs are shown in \cref{fig:BS}. In this work, as said, we employ the soft-photon approximation, which is convenient due to its factorization properties that allow an easy cancellation of IR divergencies. Furthermore, in this limit, diagrams like that in \cref{fig:BS}~(right) do not contribute.\footnote{The reason is that such an amplitude is proportional to $\epsilon_{\mu\nu\rho\sigma}p_{ijkl}^{\mu}k^{\rho}_{\gamma}$, with $k_{\gamma}\to0$.} Therefore, we only need to account for pure BS contributions like those in \cref{fig:BS}~(left), which then need to be integrated over the soft-photon energies to cancel the IR divergencies.   
The generic contribution can be expressed as
  \begin{equation}
    |\M{}{}|^2 = \int_{0}^{E_c} \frac{d^3\boldsymbol{k}}{(2\pi)^3 2E_{\boldsymbol{k}}} |\M{BS}{}|^2, \qquad 
    \M{BS}{} = e\mathcal{M}^{\textrm{LO}}\sum_f \mathcal{Q}_f \frac{p_f\cdot \varepsilon_{\gamma}^*}{p_f\cdot k_{\gamma}} +\mathcal{O}(k_{\gamma})
  \end{equation}
where $E_{\boldsymbol{k}}^2 = \boldsymbol{k}_{\gamma}^2 -m_{\gamma}^2$, $\mathcal{Q}_f$ stands for the lepton charge (we employ an IR-mass regularization) and sum over photon polarizations is implicit. The chosen $E_c$ is related to the four-lepton invariant mass as we shall see, a parameter that is closely related to the experimental setup. Summarizing, the NLO contribution can be expressed as 
  \begin{align}\label{eq:genBS}
    |\M{SF}{}|^2 ={}& e^2 |\M{LO}{}|^2  \Big( 2I(p_1,p_2) + 2I(p_3,p_4) +2I(p_1,p_4) +2I(p_2,p_3) -2I(p_2,p_4) -2I(p_1,p_3) \nonumber\\
              &  -I(p_1,p_1) -I(p_2,p_2) -I(p_3,p_3) -I(p_4,p_4)  \Big),
  \end{align}
where $I(p_i,p_j) = (p_i\cdot p_j) J(p_i,p_j)$, with the latter given as~\cite{Kampf:2005tz}
  \begin{equation}\label{eq:bsint}
  J(p_i,p_j) = \int_{0}^{E_{c}} \frac{d^3\boldsymbol{k}}{(2\pi)^3 2E_{\boldsymbol{k}}}\frac{1}{(p\cdot k_{\gamma})(p'\cdot k_{\gamma})} 
           \qquad \\
        =  \frac{1}{2(2\pi)^2}\int_0^1 \frac{dx}{q^2}\left[ \ln\left(\frac{4E_c^2}{m_{\gamma}^2}\right) 
            + \frac{q^0}{\boldsymbol{q}} \ln\left(\frac{q^0-\boldsymbol{q}}{q^0+\boldsymbol{q}}\right) \right],
  \end{equation}
with $q = xp_i + (1-x)p_j$ and $\mathcal{O}(m_{\gamma})$ terms neglected. The general integral has been solved in Ref.~\cite{tHooft:1978jhc} and is given in \cref{sec:bsint}. For identical momenta, integration is trivial and yields
  \begin{equation}\label{eq:BSeq}
    I(p_i,p_i)
    = \frac{1}{4\pi^2}\left[ \ln\left(\frac{2E_c}{m_{\gamma}}\right) 
       + \frac{1+\delta_{i,jkl}}{2\lambda_{i,jkl}} \ln\left(\frac{\Omega_i^-}{\Omega_i^+}\right)\right].
  \end{equation}
Note the difference with respect to Ref.~\cite{Barker:2002ib} that seems to assign $\boldsymbol{p}_i/p_i^0 \to \lambda_{ii}$, which is bizarre since---according to their definitions---$\lambda_{ii}\to 0$, reducing the second term to $-1$. Given the numerical differences that we anticipate, this could be a source of them unless it corresponds to a typo. For two different momenta we find, in good agreement with Ref.~\cite{Barker:2002ib}, 
  \begin{multline}\label{eq:Ipipj}
    I(p_i,p_j) =  \frac{z_{ij}}{8\pi^2\lambda_{i,j}} 
    \Bigg[  
    \ln\left(\frac{z_{i,j}+\lambda_{i,j}}{z_{i,j}-\lambda_{i,j}}\right)\ln\left(\frac{2E_c}{m_{\gamma}}\right) +
    \frac{1}{4}\ln^2\left(\frac{\Omega_i^-}{\Omega_i^+}\right) -  \frac{1}{4}\ln^2\left(\frac{\Omega_j^-}{\Omega_j^+}\right) \\
    + \operatorname{Li}_2 \left( 1-\frac{\Upsilon_{ij}\Omega_i^+}{x_{ij}\lambda_{ij}}\right) 
    + \operatorname{Li}_2 \left( 1-\frac{\Upsilon_{ij}\Omega_i^-}{x_{ij}\lambda_{ij}}\right) 
    - \operatorname{Li}_2 \left( 1-\frac{\Upsilon_{ij}\Omega_j^+}{x_{ij}\lambda_{ij}}\right) 
    - \operatorname{Li}_2 \left( 1-\frac{\Upsilon_{ij}\Omega_j^-}{x_{ij}\lambda_{ij}}\right)
    \Bigg],
  \end{multline}
where the variables above have been defined in \cref{sec:bsint}.

%%%%%%%%%%%%%%%%%%%%%%%%%%%%%%%%%%%%%%%%%%%%%%%%%%%%%%%%%%%%%%%%%
%%%%%%%%%%%%%     NLO: VACUUM POLARIZATION     %%%%%%%%%%%%%%%%%%
%%%%%%%%%%%%%%%%%%%%%%%%%%%%%%%%%%%%%%%%%%%%%%%%%%%%%%%%%%%%%%%%%
\subsection{\large{Vacuum polarization}}\label{sec:VP}

\begin{figure}[t]
    \centering
    \includegraphics[width=0.7\textwidth]{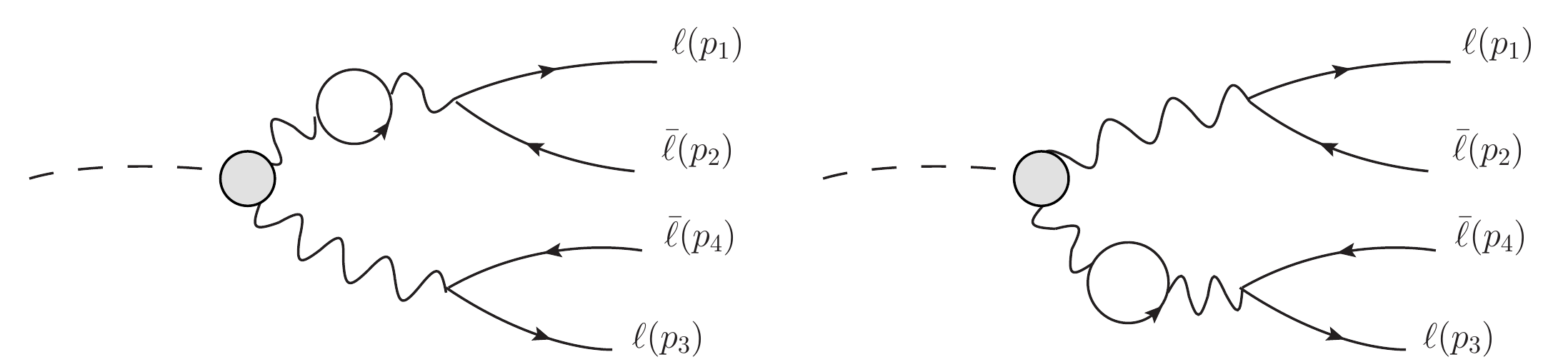}
    \caption{Contributions to the VP. Counterterms and exchange diagrams are implied.\label{fig:vpol}}
  \end{figure}

The vacuum polarization (VP), shown in \cref{fig:vpol}, induces additional contributions which simply multiply the LO ones as follows:
  \begin{align}
     \M{VP}{} = \M{LO}{D}\left[ \widehat{\Pi}(s_{12}) +\widehat{\Pi}(s_{34}) \right]
              + \left( \M{LO}{E}\left[ \widehat{\Pi}(s_{14}) +\widehat{\Pi}(s_{32})\right] \right),
  \end{align}
where $\widehat{\Pi}(q^2)$ represents the renormalized vacuum polarization. This implies a summation over the different lepton and scalar species\footnote{We consider $\ell = e, \mu$ and $s = \pi^{\pm}$ contributions. The latter was not included in Ref.~\cite{Barker:2002ib}, but is added here given that $m_{\mu}\sim m_{\pi}$. In any case, this does not produce a large impact. For the $\eta'$ case, an appropriate description for the hadronic vacuum polarization would be required though.} whose individual contributions read
  \begin{equation}
    \begin{array}{l}
      \displaystyle\widehat{\Pi}_{\ell}(q^2) =    
       -\frac{\alpha}{3\pi}\left(   \frac{8}{3} - \sigma^2 +\frac{1}{2}(3-\sigma^2)(\sigma L)  \right) \\
      \displaystyle\widehat{\Pi}_{s}(q^2) = 
        -\frac{\alpha}{6\pi}\left(   \frac{1}{3} + \sigma^2 + \frac{\sigma^2}{2}(\sigma L)  \right)  
    \end{array}
    \quad\sigma L =
    \begin{cases}
      \sigma \ln \left( \frac{\sigma-1}{1+\sigma}  \right)  & q^2 < 0 \\
      -2(1+ (3\sigma^2)^{-1})                               & q^2\to 0 \\
      -2\rho \tan^{-1}(\rho^{-1})                           & 0 < q^2 < 4m_{\ell,s}^2 \\
      \sigma\left[ \ln\left(\frac{1-\sigma}{\sigma+1}\right) + i\pi  \right]    & q^2>4m_{\ell,s}^2 \\
    \end{cases}
  \end{equation}
where $\sigma^2=-\rho^2 = 1-4m_{\ell,s}^2/q^2$. 
This produces the following terms for the NLO contributions according to the notation in \cref{eq:nlogen}: 
  \begin{align}\label{eq:vpde} \!
    \operatorname{Dir}^{\textrm{NLO}} = |\M{LO}{D}|^2 2\operatorname{Re}\left( \widehat{\Pi}(s_{12}) +\widehat{\Pi}(s_{34}) \right), \
    \operatorname{Int}^{\textrm{NLO}}_{D} = 
      2\operatorname{Re} \left[ \M{LO}{D}\M{LO*}{E}\left( \widehat{\Pi}(s_{12}) +\widehat{\Pi}(s_{34}) \right) \right].
  \end{align}
%The results above are in agreement with respect to those in Ref.~\cite{Barker:2002ib}, ecxcept that is ambiguous in their NLO definition, $2\operatorname{Re} \M{VP}{}\M{LO*}{} = |\M{LO}{}|^2\sum_g\sum_{l_g}2\operatorname{Re}\widehat{\Pi}_{l_g}(s_{ij})$, how this applies to channels with identical leptons, especially regarding Int$^{\textrm{NLO}}_{D,E}$ corrections---similar applies to the next section.
The results above are in agreement with respect to those in Ref. [5], up to ambiguities in their NLO definition [$2\operatorname{Re} \M{VP}{}\M{LO*}{} = |\M{LO}{}|^2\sum_g\sum_{l_g}2\operatorname{Re}\widehat{\Pi}_{l_g}(s_{ij})$] concerning channels with identical leptons. This comment applies as well to the next section.

%%%%%%%%%%%%%%%%%%%%%%%%%%%%%%%%%%%%%%%%%%%%%%%%%%%%%%%%%%%%%%%%%
%%%%%%%%%%%%%%%     NLO: VERTEX FUNCTIONS     %%%%%%%%%%%%%%%%%%%
%%%%%%%%%%%%%%%%%%%%%%%%%%%%%%%%%%%%%%%%%%%%%%%%%%%%%%%%%%%%%%%%%
\subsection{\large{Vertex corrections}}\label{sec:vertex}

\begin{figure}[t]\centering
    \includegraphics[width=0.7\textwidth]{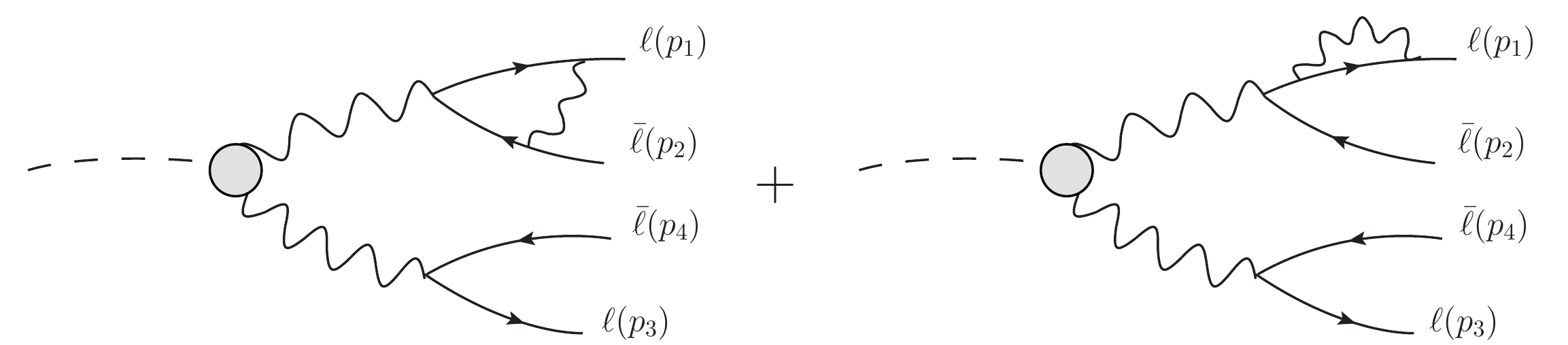}
    \caption{Contributions to the vertex corrections. Additional 
    insertion over vertices and lepton legs as well as counterterms 
    and exchange diagrams are implied.\label{fig:vertex}
    }
  \end{figure}

The vertex corrections (including lepton self-energies as usual) are shown in \cref{fig:vertex} and amount, in general, to replace the photon vertex as 
  \begin{equation}\label{eq:vertex}
    \gamma^{\mu} \to \gamma^{\mu}F_1(q^2) + \frac{i\sigma^{\mu\lambda}}{2m_{\ell}}q_{\lambda}F_2(q^2) = 
    \gamma^{\mu}(F_1(q^2)+F_2(q^2)) -\frac{\bar{q}^{\mu}}{2m_{\ell}}F_2(q^2),
  \end{equation}
where $q=p_{\ell}+ p_{\bar{\ell}}$ and $\bar{q}=p_{\ell}-p_{\bar{\ell}}$. At LO, $F_{1(2)}(q^2)=1(0)$, whereas the NLO contributions read
  \begin{align}
    \delta F_1(s) =  \frac{\alpha}{\pi} &
    \Bigg[ \left( 1+ \frac{1+\sigma^2}{2\sigma}
           \left[ \ln\left(\frac{1-\sigma}{1+\sigma}\right) +i\pi \right] \right)
           \ln\left(\frac{m_{\ell}}{m_{\gamma}}\right) -1 
           - \frac{1+2\sigma^2}{4\sigma}\left[ \ln\left(\frac{1-\sigma}{1+\sigma}\right) +i\pi \right]  \nonumber \\
    & -\frac{1+\sigma^2}{2\sigma}\Bigg( \frac{1}{4} \ln^2\left(\frac{1-\sigma}{1+\sigma}\right) 
    +\operatorname{Li}_2\left(\frac{2\sigma}{1+\sigma}\right)  -\frac{\pi^2}{2} +\frac{i\pi}{2}\ln\left(\frac{1-\sigma^2}{4\sigma^2}\right)  \Bigg)   
    \Bigg], \\
    \delta F_2(s) =   \frac{\alpha}{\pi} & \frac{1-\sigma^2}{4\sigma}\left[ \ln\left(\frac{1-\sigma}{1+\sigma}\right) +i\pi \right]; 
  \end{align}
with $\sigma=\lambda_{ij}$ for $s=s_{ij}$ and are in good agreement with the results in Ref.~\cite{Barker:2002ib}.
As a consequence, the correction due to $F_{1}$ factorizes and reduces to that in \cref{eq:vpde} upon the $\widehat{\Pi}(s_{ij}) \to \delta F_{1}(s_{ij})$ replacement. It is easy to see from \cref{eq:genBS} that IR divergencies in Dir$^{\textrm{NLO}}$ arising from $F_1$ cancel those of $2I(p_1,p_2) + 2I(p_3,p_4) - \sum_i I(p_i,p_i)$ terms---similarly, for Int$^{\textrm{NLO}}_D$, they cancel half of them. For the case of $F_{2}$, factorization in the form of Ref.~\cite{Barker:2002ib} is not obvious.\footnote{Particularly, they claim that it reduces to $|\M{LO}{}|^2 2\operatorname{Re}F_2(s_{ij}) 2(2+y_{ij}^2-\lambda_{ij}^2)^{-1}$.} Indeed, we find that 
  \begin{align}
    \operatorname{Dir}^{\textrm{NLO}} = & \frac{e^8 |F_{P\gamma\gamma}(s_{12},s_{34})|^2}{x_{12}x_{34}}\lambda^2 
      \Big(  \left(2 - (\lambda_{34}^2 - y_{34}^2)\right)2\operatorname{Re}F_2(s_{12})  + (12)\leftrightarrow(34) \Big) \\
    \operatorname{Int}_D^{\textrm{NLO}}  = & 
     \operatorname{Re} \frac{e^8 F_{P\gamma\gamma}(s_{12},s_{34})F^*_{P\gamma\gamma}(s_{14},s_{32})}{8 x_{12}x_{34}x_{14}x_{32}}\lambda^2 
      F_2(s_{12})\Big[ 2\Xi^2 +
      w^2(2+y_{12}y_{34}-y_{12}^2)(y_{34}^2-\lambda_{34}^2) \nonumber\\ & 
      + 4x_{12}(1-z)(y_{12}^2-\lambda_{12}^2)  
      + \Xi\left( 8\eta^2 - zy_{34}(y_{12}+y_{34}) +4x_{12}\lambda_{12}^2\right) \Big] + (12)\leftrightarrow(34) .
  \end{align}
which only reduces to the result in Ref.~\cite{Barker:2002ib}, for the direct term, after $\phi$ integration---a connection which is unclear if identical leptons appear. In any case, the differences for the integrated decay width would not exist for direct (and exchange) terms and are irrelevant for interference terms.

%%%%%%%%%%%%%%%%%%%%%%%%%%%%%%%%%%%%%%%%%%%%%%%%%%%%%%%%%%%%%%%%%
%%%%%%%%%%%%%%%%%%%%     NLO: 3POINT     %%%%%%%%%%%%%%%%%%%%%%%%
%%%%%%%%%%%%%%%%%%%%%%%%%%%%%%%%%%%%%%%%%%%%%%%%%%%%%%%%%%%%%%%%%
\subsection{\large{Three-point amplitudes}}\label{sec:3p}

The three-point amplitudes are the first set of RCs that were not computed in Ref.~\cite{Barker:2002ib}, which contains four different diagrams (and 4 additional exchange diagrams for identical leptons).
  \begin{figure}
    \includegraphics[width=\textwidth]{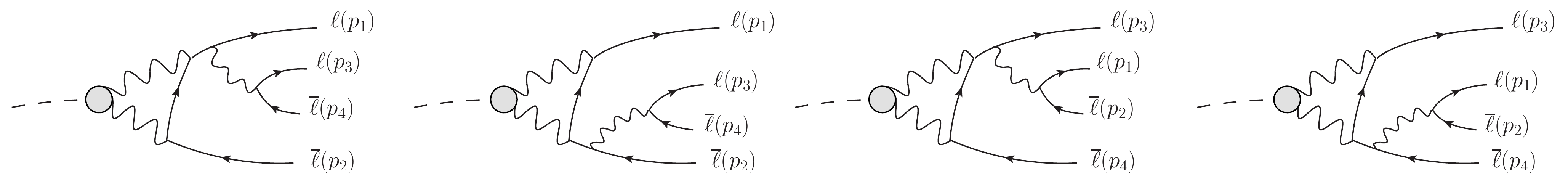}
    \caption{The three-point amplitudes for different lepton species 
             (noted in the text as 1D, 2D, 3D and 4D). If identical 
             leptons are present, exchange diagrams (eg. 
             $\overline{\ell}_2 \leftrightarrow \overline{\ell}_4$) 
             arise.\label{fig:3P}
    }
  \end{figure}
Such a contribution is UV divergent for a constant TFF and would require the same counterterm appearing in $P\to\bar{\ell}\ell$ decays~\cite{Savage:1992ac,Masjuan:2015cjl,Vasko:2011pi,Husek:2014tna}. In the following, we consider a nonconstant TFF that can be decomposed into massive-like photon propagators (see \cref{sec:TFF}) and refer to \cref{sec:3pchpt} for the case of a constant TFF. For identical leptons, the diagrams are shown in \cref{fig:3P} and their amplitudes read\footnote{To derive these, we made use of the equations of motion for spinors as well as the four-dimensional identity $\epsilon^{\alpha\beta\mu\nu}\gamma_{\mu}\gamma_{\rho}\gamma_{\nu}k_{\alpha}k^{\rho}= 2ik_{\alpha}k^{\rho}(g^{\alpha}_{\rho}\gamma^{\beta} -g^{\beta}_{\rho}\gamma^{\alpha})\gamma^5$ (note that $\epsilon^{0123}=+1$). For comments on regularization, see Ref.~\cite{Husek:2014tna}.}
  \begin{align}
    i\MM{3P}{1D} ={}& \mathcal{C}_{\textrm{3P}} \int\frac{d^4k}{(2\pi)^4}  
    \frac{[\overline{u}_1 \gamma^{\lambda}(\slashed{p}_{134}+m_a)\Gamma_{\textrm{3P}} v_2][\overline{u}_3\gamma_{\lambda}v_4]}{k^2(k+P)^2((k+p_2)^2-m_a^2)} \frac{F_{P\gamma\gamma}(k^2,(k+P)^2)}{p_{34}^2 (p_{134}^2-m_a^2)}, \\
    i\MM{3P}{2D} ={}&  \mathcal{C}_{\textrm{3P}}  \int\frac{d^4k}{(2\pi)^4}  
    \frac{[\overline{u}_1 \Gamma_{\textrm{3P}} (-\slashed{p}_{234}+m_a)\gamma^{\lambda}v_2][\overline{u}_3\gamma_{\lambda}v_4]}{k^2(k+P)^2((k+p_1)^2-m_a^2)} \frac{F_{P\gamma\gamma}(k^2,(k+P)^2)}{p_{34}^2 (p_{234}^2-m_a^2)}, 
  \end{align}
with $\MM{3P}{3D,4D}$ given upon $(12)\leftrightarrow(34)$ replacement (exchange terms would have a relative sign and exchanged $2\leftrightarrow4$ subscripts). In the expressions above, $\mathcal{C}_{\textrm{3P}}=e^4\left(\frac{i}{16\pi^2}\right)^{-1}\frac{\alpha}{2\pi}$ and $\Gamma_{\textrm{3P}}=(k^2\slashed{P} - (k\cdot P) \slashed{k})\gamma^5$ have been introduced. In addition, it is easy to show using the properties of charge conjugation that $\MM{3P}{2D} =  \MM{3P}{1D}(p_1\leftrightarrow p_2)$, which is related to $y_{12}\to-y_{12}$ and $\phi\to\phi+\pi$ replacements. Since these are symmetries of $\M{LO}{D}$, it is possible to obtain all the Dir$^{\textrm{NLO}}$ contributions from only one of them. Introducing the loop integrals and associated functions\footnote{We use the conventions for the Passarino-Veltman functions in~\cite{Hahn:2006qw}. As explained in \cref{sec:TFF}, the TFF effect can be reduced to a sum of massive-photon propagators, for which we introduce photon masses $M_{V_{1,2}}$ which should be summed over---see \cref{sec:TFF} for further details. For a constant TFF, $M_{V_i} \to 0$.}
  \begin{align}
    \mathcal{I}_1 ={}&  B_0(p_{134}^2,M_{V_2}^2,m_a^2) + M_{V_1}^2 C_0(M^2,p_{134}^2,m_a^2,M_{V_1}^2,M_{V_2}^2,m_a^2) \\
    \mathcal{I}_2^a ={}&  C_{00} + M^2C_{11} + (p_2 \cdot P)C_{12}, \quad \mathcal{I}_2^b =  M^2C_{12} +  (p_2 \cdot P)C_{22}, 
  \end{align}
where $C_{\mu\nu}=C_{\mu\nu}(M^2,p_{134}^2,m_a^2,M_{V_1}^2,M_{V_2}^2,m_a^2)$, the first contribution to Dir$^{\textrm{NLO}}$ reads 
  \begin{multline}\label{eq:nlo3p}
    2\operatorname{Re}\M{LO*}{D}\MM{3P}{1D} = 2\operatorname{Re}
    \frac{e^8 F_{P\gamma\gamma}^*(s_{12},s_{34})}{x_{34}} \frac{\alpha}{4\pi}\Bigg[ 
    \left( \mathcal{I}_1 - \mathcal{I}_2^a \right)	
     \Bigg( 4\lambda y_{12}(2+y_{34}^2-\lambda_{34}^2) -\frac{\lambda zy_{34}\Xi}{x_{12}x_{34}} \\
              +\frac{2M^2\lambda^2(1-\lambda_{12}^2)(2+y_{34}^2-\lambda_{34}^2)}{p_{134}^2 - m_a^2}  \Bigg)          
    -  \mathcal{I}_2^b\left( \frac{M^2\lambda^2(1-\lambda_{12}^2)(2+y_{34}^2-\lambda_{34}^2)}{p_{134}^2 - m_a^2} \right)  \Bigg],
  \end{multline}
where $p_{134}^2 -m_a^2 = (M^2-s_{12}+s_{34}+M^2\lambda y_{12})/2 $ and $P\cdot p_2 = (M^2+s_{12}-s_{34}-M^2\lambda y_{12})/2$. The remaining contributions can be obtained then upon the appropriate $\{s_{ij}, y_{ij}, \phi\}$ replacements. Regarding contributions of the Int$^{\textrm{NLO}}_{D}$ kind, only $p_{1,2}\leftrightarrow p_{3,4}$ is a symmetry for $\M{LO}{E}$, and two terms must be computed. These can be expressed as  
  \begin{align}
    2\operatorname{Re}\M{LO*}{E}\MM{3P}{1D} ={}& 
    -2\operatorname{Re}\frac{e^8 F_{P\gamma\gamma}^*(s_{14},s_{32})}{s_{14}s_{32}(p_{134}^2-m^2)} \frac{\alpha}{2\pi}
     \left( (\mathcal{I}_1 - \mathcal{I}_2^a)\operatorname{tr}(1D)\vert_P - (\mathcal{I}_2^b) \operatorname{tr}(1D)\vert_{p_2} \right),\label{eq:3pint1}\\
     2\operatorname{Re}\M{LO*}{E}\MM{3P}{2D} ={}& 
     -2\operatorname{Re}\frac{e^8 F_{P\gamma\gamma}^*(s_{14},s_{32})}{s_{14}s_{32}(p_{234}^2-m^2)} \frac{\alpha}{2\pi}
     \left( (\mathcal{I}'_1 - \mathcal{I'}_2^a)\operatorname{tr}(2D)\vert_P - (\mathcal{I'}_2^b) \operatorname{tr}(2D)\vert_{p_1} \right),\label{eq:3pint2}
  \end{align}
with the remaining ones obtained upon the $(12)\leftrightarrow(34)$ replacement. The meaning for $\mathcal{I}_{1,2}^{(a,b)}$ is identical as in the previous case and the primed ones amount to $p_2\to p_1$ replacement. In addition, the following traces have been introduced
  \begin{align}
    \operatorname{tr}(1D)\vert_l ={}& i \operatorname{tr} 
    \left[ (\slashed{p}_1+m)\gamma^{\lambda}(\slashed{p}_{134}+m)\slashed{l}\gamma^5(\slashed{p}_2-m)\gamma^{\sigma}(\slashed{p}_3+m)\gamma_{\lambda}(\slashed{p}_4-m)\gamma^{\nu} \right] \epsilon_{\mu\nu\rho\sigma}p_{14}^{\mu}p^{\rho}_{32},\\
    \operatorname{tr}(2D)\vert_l ={}& i \operatorname{tr} 
    \left[ (\slashed{p}_1+m)\slashed{l}\gamma^5(-\slashed{p}_{234}+m)\gamma^{\lambda}(\slashed{p}_2-m)\gamma^{\sigma}(\slashed{p}_3+m)\gamma_{\lambda}(\slashed{p}_4-m)\gamma^{\nu} \right] \epsilon_{\mu\nu\rho\sigma}p_{14}^{\mu}p^{\rho}_{32}.
  \end{align}
The resulting expressions are long but otherwise straightforward to evaluate with {\texttt{Feyncalc}}~\cite{Mertig:1990an,Shtabovenko:2016sxi}.

%%%%%%%%%%%%%%%%%%%%%%%%%%%%%%%%%%%%%%%%%%%%%%%%%%%%%%%%%%%%%%%%%
%%%%%%%%%%%%%%%%%%%%     NLO: 4POINT     %%%%%%%%%%%%%%%%%%%%%%%%
%%%%%%%%%%%%%%%%%%%%%%%%%%%%%%%%%%%%%%%%%%%%%%%%%%%%%%%%%%%%%%%%%
\subsection{\large{Four-point amplitudes}}\label{sec:4p}

\begin{figure}\centering
    \includegraphics[width=0.7\textwidth]{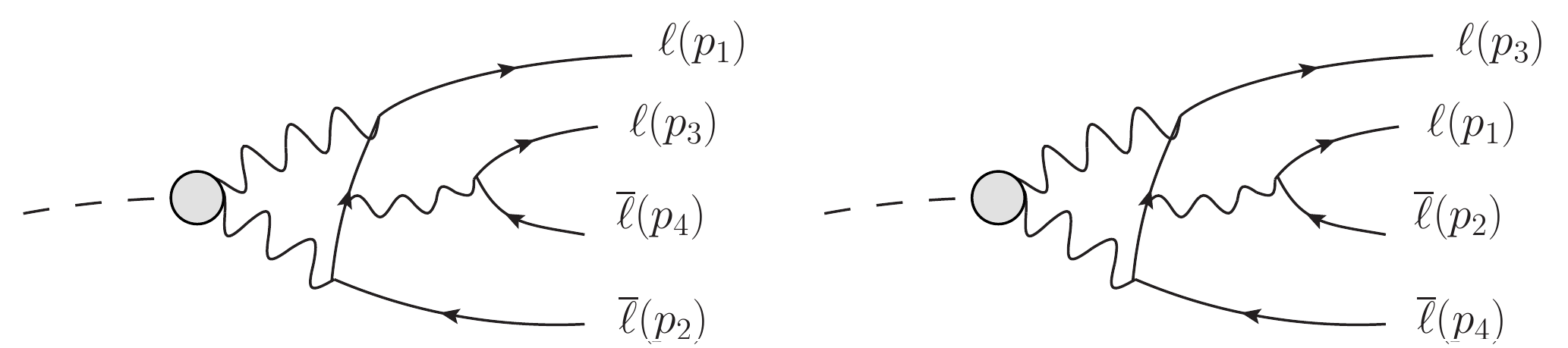}
    \caption{The four-point amplitudes for different lepton species (noted as 1D and 2D). Exchange diagrams arise for identical leptons.\label{fig:4p}}
  \end{figure}

The four-point amplitudes were not calculated in Ref.~\cite{Barker:2002ib} either and amount to a total of two contributions (another two appear for identical leptons) which are shown in \cref{fig:4p}.
The first amplitude can be expressed as\footnote{Again, we use similar manipulations as in the previous section.}
  \begin{gather}
    i\mathcal{M}_{\textrm{1D}}^{\textrm{4P}}  = \mathcal{C}_{\textrm{4P}}   \int \frac{d^4k}{(2\pi)^4}
    \frac{ \big[ \overline{u}_1 \Gamma^{\lambda}_{\textrm{4P}} v_2\big] \big[\bar{u}_3\gamma_{\lambda}v_4 \big] F_{P\gamma\gamma}(k^2,(k+P)^2)}
         {k^2 [(k+p_1)^2-m_a^2] [(k+p_{134})^2-m_a^2](k+P)^2} \frac{1}{s_{34}},\\
    \Gamma^{\lambda}_{\textrm{4P}} = 2i\left( k^{\lambda}(k+P)^2\slashed{k} - (k+P)^{\lambda}k^2(\slashed{k}+\slashed{P})  \right)\gamma^5 
    +2\epsilon_{\mu\nu\rho\sigma}k^{\mu}P^{\rho}\left(  p_1^{\nu}\gamma^{\lambda}(\slashed{k}+\slashed{P})\gamma^{\sigma} 
    + p_2^{\nu}\gamma^{\sigma}\slashed{k}\gamma^{\lambda}  \right), \label{eq:4pdec} \nonumber
  \end{gather}
with $\mathcal{C}_{\textrm{4P}} =  ie^4 \left(\frac{i}{16\pi^2}\right)^{-1} \frac{\alpha}{4\pi}$. The $\MM{4P}{2D}$ amounts to exchanging $(12)\leftrightarrow(34)$ subscripts. Again, the standard reduction into Passarino-Veltman function can be performed and equations of motion used to simplify expressions. This way, we can express the whole result for Dir$^{\textrm{NLO}}$ as
  \begin{equation}\label{eq:nlo4p}
     2\operatorname{Re}\M{LO*}{D(E)}\MM{4P}{1D} 
     = \mp 2e^8\operatorname{Re} \frac{ F^*_{P\gamma^*\gamma^*}(s_{12(14)},s_{34(32)}) }{ s_{12}s_{34}^2 (s_{14}s_{34}s_{32}) } 
     \frac{\alpha}{4\pi}\left( [...]_1 + [...]_2 + [...]_3 \right),
  \end{equation}
plus additional $(12)\leftrightarrow(34)$ terms, where $[...]_i$ stand for
  \begin{align}
    \epsilon_{\mu\nu\rho\sigma}p_{12}^{\mu}p_{34}^{\rho} \times{}& 
    \operatorname{tr}(\slashed{p}_1+m_a)\Gamma^{\lambda(i)}_{\textrm{4P}}(\slashed{p}_2-m_a)\gamma^{\nu}  \times
    \operatorname{tr}(\slashed{p}_3+m_b)\gamma_{\lambda}(\slashed{p}_4-m_b)\gamma^{\sigma}, \nonumber\\
        \epsilon_{\mu\nu\rho\sigma}p_{14}^{\mu}p_{32}^{\rho} \times{}& 
    \operatorname{tr}(\slashed{p}_1+m_a)\Gamma^{\lambda(i)}_{\textrm{4P}}(\slashed{p}_2-m_a)\gamma^{\sigma}
        (\slashed{p}_3+m_b)\gamma_{\lambda}(\slashed{p}_4-m_b)\gamma^{\nu},
  \end{align}
for $\textrm{Dir}^{\textrm{NLO}}$ and $\textrm{Int}_D^{\textrm{NLO}}$, respectively, and with $\Gamma^{\lambda}_{\textrm{4P}} =  \Gamma^{\lambda(1)}_{\textrm{4P}} +  \Gamma^{\lambda(2)}_{\textrm{4P}} +  \Gamma^{\lambda(3)}_{\textrm{4P}}$ defined below
  \begin{align}
    \Gamma^{\lambda(1)}_{\textrm{4P}} ={}& 
      2i\Big[ \gamma^{\lambda}C_{00} +m_ap_1^{\lambda}\left( C_{11} +2C_{12} +C_{22}\right) 
      +p_1^{\lambda}\slashed{p}_{34}\left(C_{12} +C_{22}\right)\Big]\gamma^5 +2iM_{V_2}^2\Big[ \gamma^{\lambda}D_{00} \nonumber \\ &
      + m_ap_1^{\lambda}\left( D_{11} +2D_{12} +D_{22} +3D_{13} +3D_{23} +2D_{33}\right) 
      +p_1^{\lambda}\slashed{p}_{34}\Big(D_{12} +D_{13} +D_{22} \nonumber \\ & 
      +2D_{23} +D_{33}\Big) +m_ap_2^{\lambda}\left( D_{13} +D_{23} +2D_{33}\right) 
      +p_2^{\lambda}\slashed{p}_{34}\left( D_{23} +D_{33}\right)\Big]\gamma^5 - (p_1\to p_2), \\
    \Gamma^{\lambda(2)}_{\textrm{4P}} ={}& 
      2\epsilon_{\mu\nu\rho\sigma}P^{\rho}p_1^{\nu}\gamma^{\lambda}\gamma^{\mu}\gamma^{\sigma}D_{00} + 
      2\epsilon_{\mu\nu\rho\sigma}p_{34}^{\mu}p_1^{\nu}p_2^{\rho}\Big( \gamma^{\lambda}\slashed{p}_1\gamma^{\sigma}
      \left( D_{12} +D_{22} +D_{23} +D_{2}\right) \nonumber\\ &
      +m_a\gamma^{\lambda}\gamma^{\sigma}\left( D_{23} +D_2\right) + \gamma^{\lambda}\slashed{p}_{34}\gamma^{\sigma}
      \left( D_{22} +D_{23} +D_2\right) \Big), \\
    \Gamma^{\lambda(3)}_{\textrm{4P}} ={}& 
      2\epsilon_{\mu\nu\rho\sigma}P^{\rho}p_2^{\nu}\gamma^{\sigma}\gamma^{\mu}\gamma^{\lambda}D_{00} + 
      2\epsilon_{\mu\nu\rho\sigma}p_{34}^{\mu}p_1^{\nu}p_2^{\rho}\Big( \gamma^{\sigma}\slashed{p}_2\gamma^{\lambda}D_{13} 
      + \gamma^{\sigma}\slashed{p}_{34}\gamma^{\lambda}\left( D_{13} +D_{12}\right)\nonumber\\ &
      -m_a\gamma^{\sigma}\gamma^{\lambda}\left( D_{11} +D_{12} +D_{13}\right)  \Big).
  \end{align}
Once more, standard Passarino-Veltman $D_{\mu\nu} = D_{\mu\nu}(m_a^2,s_{34},m_a^2,M^2,p_{134}^2,p_{234}^2,M_{V_{1}}^2,m_a^2,m_a^2,M_{V_{2}}^2)$ and $C_{\mu\nu} = C_{\mu\nu}(m_a^2,s_{34},p_{134}^2,M_{V_{1}}^2,m_a^2,m_a^2)$ functions have been introduced---see comments in \cref{sec:3p}.  The traces can be computed easily with {\texttt{Feyncalc}}.

%%%%%%%%%%%%%%%%%%%%%%%%%%%%%%%%%%%%%%%%%%%%%%%%%%%%%%%%%%%%%%%%%
%%%%%%%%%%%%%%%%%%%%     NLO: 5POINT     %%%%%%%%%%%%%%%%%%%%%%%%
%%%%%%%%%%%%%%%%%%%%%%%%%%%%%%%%%%%%%%%%%%%%%%%%%%%%%%%%%%%%%%%%%
\subsection{\large{Five-point amplitudes}}\label{sec:5p}

\begin{figure}
    \includegraphics[width=\textwidth]{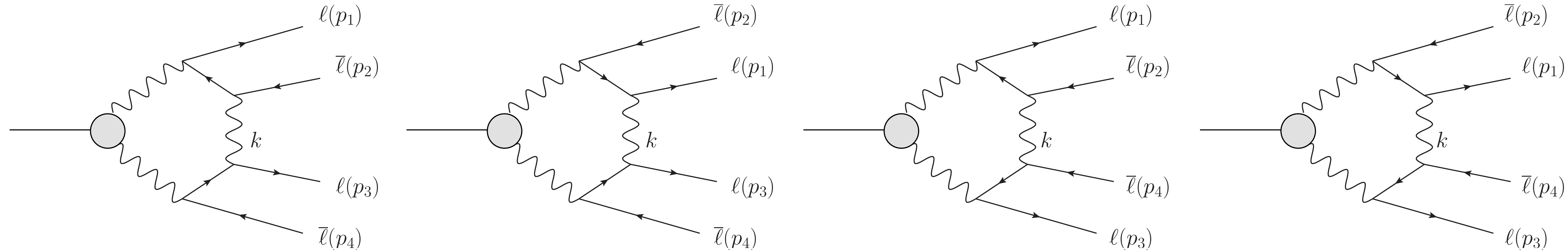}
    \caption{The pentagon diagrams contributing to the NLO corrections (noted as 1D, 2D, 3D and 4D). 
             For identical final state leptons additional exchange diagrams appear. The photon propagator 
             connecting the two letpons carries always the loop momentum $k$ in our convention.\label{fig:5p}
    }
  \end{figure}

The last set of contributions are the five-point amplitudes. There are a total of four diagrams as shown in \cref{fig:5p} 
(plus additional exchange diagrams whenever identical leptons appear). The corresponding amplitudes can be expressed, after applying the equations of motion, as 
  \begin{align}
    i\MM{5P}{1D} = & -e^6\int\frac{d^4k}{(2\pi)^4} 
       \Big( -4(p_2 \cdot p_3)(\overline{u}_1\gamma^{\nu}v_2)(\overline{u}_3\gamma^{\sigma}v_4) 
       +2k_{\alpha}\Big[ (\overline{u}_1\gamma^{\nu}\gamma^{\alpha}\slashed{p}_3v_2)(\overline{u}_3\gamma^{\sigma}v_4)  \nonumber \\&
                       -  (\overline{u}_1\gamma^{\nu}v_2)(\overline{u}_3\slashed{p}_2\gamma^{\alpha}\gamma^{\sigma}v_4) \Big] 
       + k_{\alpha}k_{\beta}(\overline{u}_1\gamma^{\nu}\gamma^{\alpha}\gamma^{\eta}v_2)(\overline{u}_3\gamma_{\eta}\gamma^{\beta}\gamma^{\sigma}v_4)
       \Big)\mathcal{C}^{\textrm{5P}}_{\nu\sigma}(p_3,p_2),\\
    i\MM{5P}{2D} = & +e^6\int\frac{d^4k}{(2\pi)^4}
       \Big( -4(p_1 \cdot p_3)(\overline{u}_1\gamma^{\nu}v_2)(\overline{u}_3\gamma^{\sigma}v_4) 
       +2k_{\alpha}\Big[ (\overline{u}_1\slashed{p}_3\gamma^{\alpha}\gamma^{\nu}v_2)(\overline{u}_3\gamma^{\sigma}v_4)  \nonumber \\ &
                       -  (\overline{u}_1\gamma^{\nu}v_2)(\overline{u}_3\slashed{p}_1\gamma^{\alpha}\gamma^{\sigma}v_4) \Big]  
       + k_{\alpha}k_{\beta}(\overline{u}_1\gamma^{\eta}\gamma^{\alpha}\gamma^{\nu}v_2)(\overline{u}_3\gamma_{\eta}\gamma^{\beta}\gamma^{\sigma}v_4)
       \Big)\mathcal{C}^{\textrm{5P}}_{\nu\sigma}(p_3,p_1) \\
    i\MM{5P}{3D} = & +e^6\int\frac{d^4k}{(2\pi)^4}
       \Big( -4(p_2 \cdot p_4)(\overline{u}_1\gamma^{\nu}v_2)(\overline{u}_3\gamma^{\sigma}v_4)  
       +2k_{\alpha}\Big[ (\overline{u}_1\gamma^{\nu}\gamma^{\alpha}\slashed{p}_4v_2)(\overline{u}_3\gamma^{\sigma}v_4) \nonumber \\&
                       -  (\overline{u}_1\gamma^{\nu}v_2)(\overline{u}_3\gamma^{\sigma}\gamma^{\alpha}\slashed{p}_2v_4) \Big]  
       + k_{\alpha}k_{\beta}(\overline{u}_1\gamma^{\nu}\gamma^{\alpha}\gamma^{\eta}v_2)(\overline{u}_3\gamma^{\sigma}\gamma^{\beta}\gamma_{\eta}v_4)
       \Big)\mathcal{C}^{\textrm{5P}}_{\nu\sigma}(p_4,p_2) \displaybreak\\
    i\MM{5P}{4D} = & -e^6\int\frac{d^4k}{(2\pi)^4}
       \Big( -4(p_1 \cdot p_4)(\overline{u}_1\gamma^{\nu}v_2)(\overline{u}_3\gamma^{\sigma}v_4) 
       +2k_{\alpha}\Big[ (\overline{u}_1\slashed{p}_4\gamma^{\alpha}\gamma^{\nu}v_2)(\overline{u}_3\gamma^{\sigma}v_4)  \nonumber \\&
                       -  (\overline{u}_1\gamma^{\nu}v_2)(\overline{u}_3\gamma^{\sigma}\gamma^{\alpha}\slashed{p}_1v_4) \Big]  
       + k_{\alpha}k_{\beta}(\overline{u}_1\gamma^{\eta}\gamma^{\alpha}\gamma^{\nu}v_2)(\overline{u}_3\gamma^{\sigma}\gamma^{\beta}\gamma_{\eta}v_4)
       \Big)\mathcal{C}^{\textrm{5P}}_{\nu\sigma}(p_4,p_1),  
  \end{align}
where 
  \begin{equation}
    \mathcal{C}^{\textrm{5P}}_{\nu\sigma}(p_i,p_j) =   
      \frac{\epsilon_{\mu\nu\rho\sigma}(p_{12}^{\mu}p_{34}^{\rho} + P^{\mu}k^{\rho}) F_{P\gamma\gamma}\left( (k -p_{12})^2, (k+ p_{34})^2 \right)} 
      {k^2 [(k+p_i^2)-m_i^2] (k+p_{34})^2 (k-p_{12})^2 [(k-p_j)^2-m_j^2]},
  \end{equation}
and is in good agreement with Ref.~\cite{Barker:2002ib}. The decomposition above is convenient, as it isolates the IR-divergent part contained in the first term. Particularly, taking $k\to 0$ and retaining only the divergent propagators in the loop integral, it is easy to show that
  \begin{align}\label{eq:5pir}
    \mathcal{M}^{\textrm{5P}}_{\textrm{D}}\Big\vert_{\textrm{IR}} = 
     &  -(\mathcal{Q}_i \cdot \mathcal{Q}_j) \  \mathcal{M}^{\textrm{LO}}  \frac{e^2}{8\pi}\frac{z_{ij}}{\lambda_{ij}} 
                     \left[ \ln\left(\frac{z_{ij} + \lambda_{ij}}{z_{ij} - \lambda_{ij}}\right) -2i\pi \right]\ln m_{\gamma},
  \end{align}
where $\mathcal{Q}_{i,j}$ denotes the charge of the particles. Comparing to \cref{eq:genBS,eq:Ipipj}, Dir$^{\textrm{NLO}}$ cancels IR divergences arising from $2I(p_1,p_4) +2I(p_2,p_3) -2I(p_2,p_4) -2I(p_1,p_3)$ terms, whereas Int$^{\textrm{NLO}}_D$ cancels the $I(p_1,p_4) +I(p_2,p_3) -I(p_2,p_4) -I(p_1,p_3)$ combination.

At this point it is important to note that the diagrams above are all related through $p_{1}\leftrightarrow p_2$ and  $p_{3}\leftrightarrow p_4$ exchanges---similar to the three-point case. Note, however, how in this case each of these changes carries a minus sign. This must actually be this way to cancel the IR divergencies as it can be observed from \cref{eq:5pir}. Again, since this is a symmetry for $\M{LO}{D}$, it is only necessary to calculate one of the contributions above---all the four Dir$^{\textrm{NLO}}$ terms will be related upon $y_{ij}\to -y_{ij}$ and $\phi\to\phi+\pi$. 
For the Int$^{\textrm{NLO}}_{D}$ terms, only the combined $p_{1,3}\leftrightarrow p_{2,4}$ exchange is a symmetry, and two terms must be computed, which implies that, for Dir$^{\textrm{NLO}}$ (but not for Int$^{\textrm{NLO}}$), the overall correction to the decay width vanishes.\footnote{This is due to charge conjugation---see for instance the comments on pg.~8 from Ref.~\cite{Fael:2016yle}.} Accounting for these simplifications, the required contributions were computed in the following way through the use of {\texttt{Feyncalc}}: first of all, we evaluated the lepton traces. Then, the resulting terms of the $p_i\cdot k$ kind were canceled against propagators as much as possible, which leaves the five-point scalar function and lower-point tensor functions.

%%%%%%%%%%%%%%%%%%%%%%%%%%%%%%%%%%%%%%%%%%%%%%%%%%%%%%%%%%%%%%%%%
%%%%%%%%%%%%%%%%%%%     NLO: SUMMARY     %%%%%%%%%%%%%%%%%%%%%%%%
%%%%%%%%%%%%%%%%%%%%%%%%%%%%%%%%%%%%%%%%%%%%%%%%%%%%%%%%%%%%%%%%%
\subsection{\large{Full NLO numerical results}}\label{sec:nlofull}

Finally, we give the numerical results that we obtain, which we carry out with the help of $\texttt{Looptools}$~\cite{Hahn:1998yk,Hahn:2006qw} for evaluating the loop integrals\footnote{As a cross-check, we computed independently the scalar five-point function, $D_0$, in terms of four-point functions using the method in Ref.~\cite{Denner:2002ii} finding good agreement. We also find agreement with higher rank five-point functions that we employed to further cross-check our results---this is not the case for {\texttt{Feyncalc}}. Note that the method in Ref.~\cite{Denner:2002ii} has the advantage of avoiding singularities in Gram determinants. } 
and the \texttt{Vegas} method in the \texttt{CUBA} library for the numerical integration.\footnote{Again, we do not need to integrate Exc or Int$^{\textrm{NLO}}_E$ terms since they contribute the same as Dir and Int$^{\textrm{NLO}}_D$, respectively. The same applies to $y_{ij}\to -y_{ij}$ and $\phi\to\phi+\pi$ related terms.} As said, the $F_1$ and five-point amplitudes contributions contain IR divergencies and are thereby combined with the appropriate BS parts to render an IR-finite result.\footnote{\color{black}We checked that the full and partial contributions were independent of the $m_{\gamma}$ parameter as they should be.} Regarding the cutoff energy for the soft photon, this is related to the four-lepton invariant mass through the 
  \begin{equation}
    E_c = \frac{M}{2}(1 - x_{4\ell}), \qquad x_{4\ell} =p_{4\ell}^2 M^{-2}
  \end{equation}
relation. In the following, we take $x_{4\ell}=0.9985$ in analogy with Ref.~\cite{Barker:2002ib}---see comments below for different cutoffs. 
Expressing $\Gamma_{4\ell} = \Gamma_{4\ell}^{\textrm{LO}} +\Gamma_{4\ell}^{\textrm{NLO}}$, we give the RC in terms of $\delta_{(\textrm{FF})} =\Gamma_{4\ell}^{\textrm{NLO}} /\Gamma_{4\ell}^{\textrm{LO}}$ in \cref{tab:NLOdelta}, where the FF subscript means that a nonconstant TFF was employed (see \cref{sec:TFF}).
  \begin{table}[tbp]\centering\small
  \begin{tabular}{cccccccc}\toprule
                   & $\pi^0\to 4e$ & $K_L\to 4e$ & $K_L\to 2e2\mu$ & $K_L\to 4\mu$ & $\eta\to 4e$ & $\eta\to 2e2\mu$ & $\eta\to 4\mu$  \\ \midrule
  $\delta $                         &$-0.1724(2)$ &$-0.2268(2)$ & $-0.0798(2)$ & $0.0669(1)$ & $-0.2306(1)$ & $-0.0864(1)$ & $0.0502(1)$ \\  
  $\delta^{\textrm{IR}}$            &$\phantom{-}0.0411(1)$  &$\phantom{-}0.0534(1)$  & $\phantom{-}0.0273(0)$  & $0.0021(0)$ & $\phantom{-}0.0543(1)$  & $\phantom{-}0.0285(1)$  & $0.0033(2)$ \\ \midrule 
  $\delta^{\textrm{Partial}}$       &$-0.1718(2)$ &$-0.2262(2)$ & $-0.0767(1)$ & $0.0704(1)$ & $-0.2301(1)$ & $-0.0836(1)$ & $0.0535(1)$ \\ 
  \cite{Barker:2002ib}              &$-0.160(2)$  &$-0.218(1)$  & $-0.066(1)$ & $0.084(1)$ & $-$ & $-$ & $-$ \\ \midrule
    %%%%%%%%%%%%%%%%%%%%%%%%%%%%%%%%%%%%%%%%%%%%%%%%%%%%%%%%%%%%%%
  $\delta_{\textrm{FF}}$ &   $-0.1727(2)$ & $-0.2345(1)$ & $-0.0842(2)$ & $0.0608(2)$ & $-0.2409(1)$ & $-0.0900(1)$ & $0.0455(2)$ \\  
  $\delta^{\textrm{IR}}_{\textrm{FF}}$ & $\phantom{-}0.0411(1)$ & $\phantom{-}0.0549(1)$ & $\phantom{-}0.0276(0)$ & $0.0022(0)$ & $\phantom{-}0.0554(1)$ & $\phantom{-}0.0288(1)$ & $0.0033(0)$ \\  \bottomrule
  \end{tabular}\caption{Results for the NLO RC expressed as $\delta =\Gamma^{\textrm{NLO}}_{4\ell}/\Gamma^{\textrm{LO}}_{4\ell}$; find details in the text.\label{tab:NLOdelta}}
  \end{table}
For details concerning individual NLO contributions, we refer to \cref{tab:NLOdetail} and \cref{tab:NLOdetailFF} for constant and $q^2$-dependent TFFs, respectively. We note that we do not ascribe any error to the TFF description, which is intended mainly to illustrate the magnitude of TFF effects against RC. 
Concerning extrapolations to different $x_{4\ell}$ values, we integrate the divergent $\ln m_{\gamma}$ terms, so that extrapolation to different cutoffs can be obtained through
  \begin{equation}
    \delta_{(\textrm{FF})}(x_{4\ell}) = \delta_{(\textrm{FF})}(0.9985) + \delta_{(\textrm{FF})}^{\textrm{IR}} \ln\left( \frac{1-x_{4\ell}}{0.0015} \right),  
  \end{equation}
with $\delta^{\textrm{IR}}_{(\textrm{FF})}$ given in \cref{tab:NLOdelta}. We stress that such result holds in the soft-photon approximation, this is, it is not meant to be used to obtain a fully inclusive ($P\to 4\ell\gamma$) decay width. 

We give as well our result without including three- and four-point contributions and for a constant TFF ($\delta^{\textrm{Partial}}$ column in \cref{tab:NLOdelta}). This compares to Ref.~\cite{Barker:2002ib} results, which we give for convenience in the fifth column from \cref{tab:NLOdelta}.\footnote{These are obtained from the results in Tables VI and VII in Ref. [5].} As a result, we find discrepancies at the $1\%$ level, which is nevertheless often of similar size as TFF effects (see \cref{tab:LO} and, especially, the $\pi^0$ case). From our results in \cref{tab:NLOdetail}, we find out that such effects can only arise from VP, $F_1$ and BS contributions; these corrections were computed analytically and we agree with all of them except for their $I(p_i,p_i)$ result which, as said, is unclear. A different source of discrepancy would be an underestimated statistical uncertainty associated to their MC simulation---in this respect, we note that we checked our results against the {\texttt{NIntegrate}} method in {\texttt{Mathematica}} for these contributions, finding an excellent agreement.

Finally, it is worth commenting about the three-point contribution when employing constant TFFs. In such case it is necessary to employ $\chi$PT, which introduces a counterterm that is connected to $P\to\bar{\ell}\ell$ decays. We find, however, that such a practice might be inappropriate for muons and including the TFF is desirable (find more details in \cref{sec:3pchpt}).

In summary, we find relevant numerical differences for the contributions calculated in Ref.~\cite{Barker:2002ib} with a non-negligible effect regarding the extraction of the TFF. Concerning the new three- and four-point loop contributions, these are small as compared to the full NLO correction, but of similar size as $F_2$ and 5P contributions. Note, however, that such considerations have to be taken with care if considering differential distributions as required in experiments.

%%%%%%%%%%%%%%%%%%%%%%%%%%%%%%%%%%%%%%%%%%%%%%%%%%%%%%%%%%%%%%%%%
%%%%%%%%%%%%%%%%%%%%%%%%%%%%%%%%%%%%%%%%%%%%%%%%%%%%%%%%%%%%%%%%%
%%%%%%%%%%%%%%%%%%%%%                     %%%%%%%%%%%%%%%%%%%%%%%
%%%%%%%%%%%%%%%%%%%%%     TFF EFFECTS     %%%%%%%%%%%%%%%%%%%%%%%
%%%%%%%%%%%%%%%%%%%%%                     %%%%%%%%%%%%%%%%%%%%%%%
%%%%%%%%%%%%%%%%%%%%%%%%%%%%%%%%%%%%%%%%%%%%%%%%%%%%%%%%%%%%%%%%%
%%%%%%%%%%%%%%%%%%%%%%%%%%%%%%%%%%%%%%%%%%%%%%%%%%%%%%%%%%%%%%%%%
\section{\large{TFF Effects}}\label{sec:TFFeff}

As said, these decays are of interest for obtaining relevant information on the TFFs; as an example, see the works in Refs.~\cite{Bijnens:1999jp,Petri:2010ea,Terschlusen:2013iqa,Escribano:2015vjz}. In the following, we comment briefly on some aspects that, we believe, could be tested at future experiments. 

Concerning the $\pi^0$, the highest double-virtual region that can be accessed, $q_1^2=q_2^2=q_{\textrm{max}}^2=m_{\pi}^2/4$, is small enough to rely on a series expansion to parametrize the TFF. Consequently, such effects would be as small as $(m_{\pi}/2\Lambda)^4$, where $\Lambda$ is expected to be the order of $M_{V_{\pi^0}}$ (see \cref{sec:TFF} and Refs.~\cite{Masjuan:2015lca,Masjuan:2015cjl,Masjuan:2017tvw}). In addition, since the process peaks at low energies, the double-virtual region is---experimentally---less populated. As a consequence, we think that only the TFF slope could be accessed experimentally. 
In this respect, the single result comes from KTeV~\cite{Abouzaid:2008cd} (with 30511 events and $0.7\%$ precision), which found a negative (yet compatible with $0$) value, in contradiction with current results (find experimental references in~\cite{Masjuan:2017tvw}), an outcome that could be due to statistics, systematics, or RC. Regarding the latter, from \cref{tab:LO}, TFF effects are of order $0.4\%$, whereas the differences found for the RC~\cite{Barker:2002ib} employed in \cite{Abouzaid:2008cd} is of $1.24\%$, 3 times larger. Of course, a differential analysis in the lines of Ref.~\cite{Abouzaid:2008cd} would be of relevance in order to draw firm conclusions. In this aspect, the NA62 Collaboration, already successful in obtaining the best measurement for the $\pi^0\to e^+e^-\gamma$~\cite{TheNA62:2016fhr}, could make advances in this direction.

Concerning the $\eta$, the larger available phase space could make the process interesting for accessing the double-virtual region. So far, the only available result is for $\eta\to 4e$ from KLOE~Collaboration~\cite{KLOE2:2011aa} (with 362 events and $10\%$ precision), which did not attempt a fit to the TFF, likely due to the low statistics. In the future, the REDTOP~Collaboration~\cite{Gatto:2016rae} could have larger statistics for all the $\eta$ channels, which would provide very interesting results---we note here that, for the electronic channel, TFF effects are of order $6\%$ (see \cref{tab:LO}), whereby the differences found with respect to RC in Ref.~\cite{Barker:2002ib} are relevant. What is more, if entering the $\eta'$ mode, the REDTOP Collaboration would undoubtedly test the double-virtual region, yet this makes necessary an appropriate description for the resonant structure, which is left for future work. 

Eventually, if the double-virtual region is accessed, this might be of interest regarding the HLbL contribution to the muon $g-2$~\cite{Masjuan:2017tvw}. Another possibility to access such a region, closely related to this process by crossing, are the $e^+e^- \to P e^+e^- $ processes, in which study of RC is postponed for future investigation.

%%%%%%%%%%%%%%%%%%%%%%%%%%%%%%%%%%%%%%%%%%%%%%%%%%%%%%%%%%%%%%%%%
%%%%%%%%%%%%%%%%%%%%%%%%%%%%%%%%%%%%%%%%%%%%%%%%%%%%%%%%%%%%%%%%%
%%%%%%%%%%%%%%%%%%%%%                     %%%%%%%%%%%%%%%%%%%%%%%
%%%%%%%%%%%%%%%%%%%%%     CONCLUSIONS     %%%%%%%%%%%%%%%%%%%%%%%
%%%%%%%%%%%%%%%%%%%%%                     %%%%%%%%%%%%%%%%%%%%%%%
%%%%%%%%%%%%%%%%%%%%%%%%%%%%%%%%%%%%%%%%%%%%%%%%%%%%%%%%%%%%%%%%%
%%%%%%%%%%%%%%%%%%%%%%%%%%%%%%%%%%%%%%%%%%%%%%%%%%%%%%%%%%%%%%%%%
\section{\large{Conclusions}}\label{sec:Conclusions}

In summary, we have revisited and completed the full NLO corrections to $P\to \bar\ell\ell \bar\ell'\ell'$ processes within the soft-photon approximation, whose full result is available in a \texttt{Mathematica} notebook upon request. As a result, we found differences of the $1\%$ order with respect to the existing ones~\cite{Barker:2002ib}---likely to be relevant for extracting information about the TFF. 

Regarding the double-virtual TFF effects, these might be accessed for the $\eta$ and $\eta'$ cases. Otherwise, it might be interesting to look at the $e^+e^-\to Pe^+e^-$ processes, which are also of relevance for testing exclusive processes in pQCD; we postpone the study of RC therein for future work. 

The authors acknowledge A.~Nyffeler for discussions regarding the TFF effects, and to A.~Kupsc and H.~Czy\.z for pointing to the $e^+e^-\to Pe^+e^-$ related process and discussions on it. P.~S.-P. is indebted to Vladyslav Shtabovenko for help with Feyncalc and Tom\'a\v{s} Husek for discussions. This work was supported by the Czech Science Foundation (grant no.~GACR 18-17224S) and by the project UNCE/SCI/013 of Charles University.

%%%%%%%%%%%%%%%%%%%%%%%%%%%%%%%%%%%%%%%%%%%%%%%%%%%%%%%%%%%%%%%%%
%%%%%%%%%%%%%%%%%%%%%%%%%%%%%%%%%%%%%%%%%%%%%%%%%%%%%%%%%%%%%%%%%
%%%%%%%%%%%%%%%%%%%%%                     %%%%%%%%%%%%%%%%%%%%%%%
%%%%%%%%%%%%%%%%%%%%%     APPENDICES      %%%%%%%%%%%%%%%%%%%%%%%
%%%%%%%%%%%%%%%%%%%%%                     %%%%%%%%%%%%%%%%%%%%%%%
%%%%%%%%%%%%%%%%%%%%%%%%%%%%%%%%%%%%%%%%%%%%%%%%%%%%%%%%%%%%%%%%%
%%%%%%%%%%%%%%%%%%%%%%%%%%%%%%%%%%%%%%%%%%%%%%%%%%%%%%%%%%%%%%%%%
\appendix

%%%%%%%%%%%%%%%%%%%%%%%%%%%%%%%%%%%%%%%%%%%%%%%%%%%%%%%%%%%%%%%%%
%%%%%%%%%%%%%%%%%% KINEMATICS & PHASE SPACE %%%%%%%%%%%%%%%%%%%%%
%%%%%%%%%%%%%%%%%%%%%%%%%%%%%%%%%%%%%%%%%%%%%%%%%%%%%%%%%%%%%%%%%
\section{\large{Kinematics and phase space}}\label{sec:kin}

\begin{figure}[t]\centering
    \includegraphics[width=0.49\textwidth]{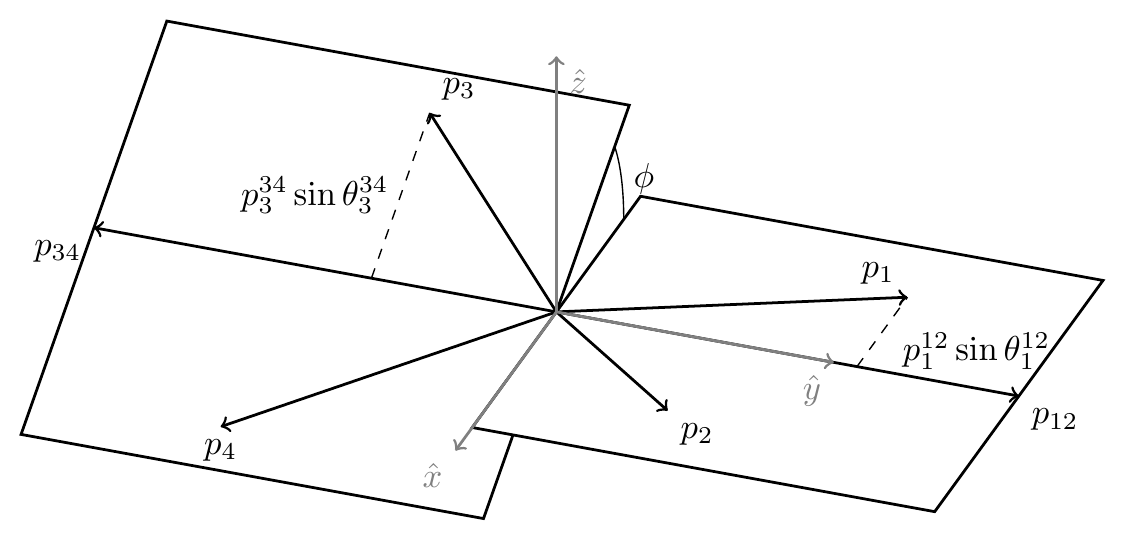}
    \includegraphics[width=0.49\textwidth]{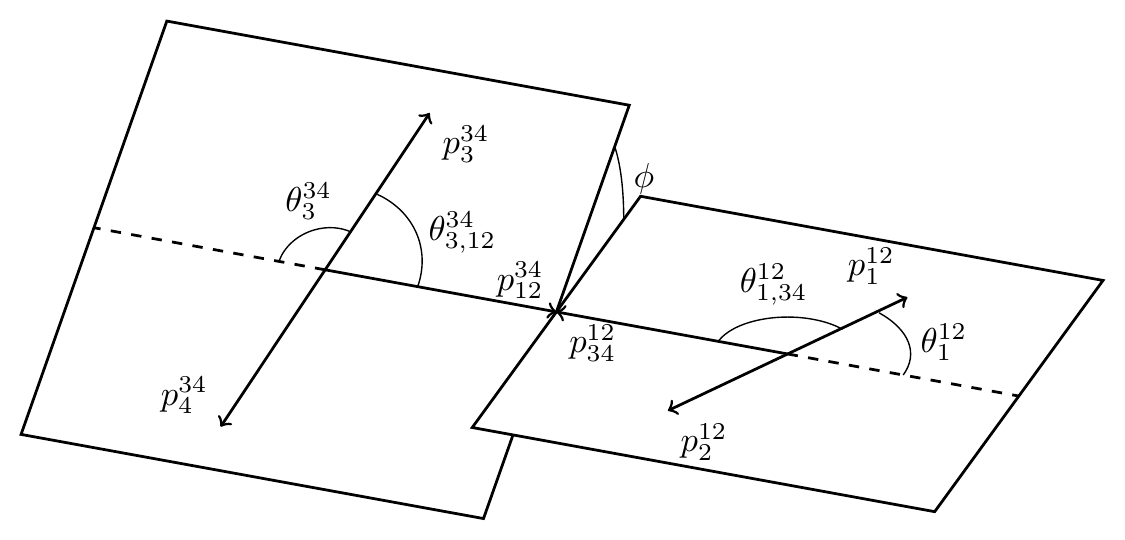}
    \caption{The left figure shows the kinematics in the parent particle rest frame. The right 
    one shows the angle of interest (e.g., in the dilepton reference frame) for phase space 
    $\theta_{1}^{12}$ and $\theta_{3}^{34}$.\label{fig:kin}}
  \end{figure}

The kinematics of the process is shown in \cref{fig:kin} (left),\footnote{Ref.~\cite{Barker:2002ib} uses opposite labeling for particles, so comparing \cref{fig:kin} and Feynman diagrams requires $p_{1(3)}\leftrightarrow p_{2(4)}$.} with momentum and mass assignment $\ell(p_1,m_a)$, $\bar{\ell}(p_2,m_a)$, $\ell'(p_3,m_b)$, $\bar{\ell}'(p_4,m_b)$.
The resulting phase space can be described sequentially in terms of a two-body decay in the parent particle rest frame to dilepton subsystems $p_1+p_2\equiv p_{12}$ and $p_3+p_4\equiv p_{34}$ with corresponding invariant masses $p_{12}^2 = s_{12}$ and $p_{34}^2 = s_{34}$ and followed by a sequential two-body decay in the corresponding subsystems' rest frames  [see \cref{fig:kin} (right), where superscripts denote the reference frame]. To see this, insert the identity as $ \int d\vec{p}_{ij}(2p_{ij}^0)^{-1} ds_{ij} \delta^{(4)}(p_{ij} - p_i -p_j)$ into the four-body phase space to obtain\footnote{For a particle decaying into $n$ particles, $d\Phi_n(P;p_1,...,p_n) = (2\pi)^4\delta^{(4)}(P-\sum_{i=1}^n p_i) \prod_{i=1}^n \frac{d\vec{p}_i}{(2\pi)^3 2E_i} $. }
  \begin{equation}
    d\Phi_4(P;p_1,p_2,p_3,p_4) = d\Phi_2(P;p_{12},p_{34})\times\frac{ds_{12}}{2\pi}d\Phi_2(p_{12};p_{1},p_{2})\times\frac{ds_{34}}{2\pi}d\Phi_2(p_{34};p_{3},p_{4}).
  \end{equation}
Before continuing, it is useful to introduce some notation, which we choose similar to Ref.~\cite{Barker:2002ib} for ease of comparison. Adopting $p_{i_1 i_2...i_n}\equiv p_{i_1}+p_{i_2}+...+p_{i_n}$, we define the following quantities, 
  \begin{equation}
   \delta_{ij} = (p_i^2 - p_j^2)/p_{ij}^2, \quad x_{i} = p_{i}^2/p_{ij}^2, 
    \quad w_{ij}^2 = 4x_ix_j, \quad z_{ij} = 2p_i\cdot p_j/p_{ij}^2 = 1-x_i-x_j,
  \end{equation}
allowing us to express the energy and momenta of a particle $p_{i(j)}$ in the $p_{ij}$ rest frame as 
  \begin{equation}\label{eq:kindefs}
    E_{i(j)}^{ij} = \sqrt{s_{ij}}(1\pm\delta_{ij})/2, \quad p_{i(j)}^{ij} = \sqrt{s_{ij}}\lambda_{ij}/2, \quad
    \lambda_{ij} = \sqrt{z_{ij}^2-w_{ij}^2}. 
  \end{equation}
In addition, whenever we use $p_{ijk}=p_{i}+p_{jk}$ configurations (or similar), we employ expressions of the kind $\lambda_{i,jk}$. Using this and conventions in \cref{fig:kin}, it is possible to express the Lorentz-invariant remaining quantities as 
  \begin{align}
    \label{eq:z}
    2 p_{12}\cdot p_{34} & = M^2 z, \\
    %%%%%%
    \label{eq:y12}
    2\bar{p}_{12}\cdot p_{34} & = \lambda M^2 y_{12} \\
    %%%%%%
    \label{eq:y34}
    2\bar{p}_{34}\cdot p_{12} & = \lambda M^2 y_{34} \\
    %%%%%%
    \label{eq:xi}
    2\bar{p}_{12}\cdot\bar{p}_{34} & = 
     M^2(  zy_{12}y_{34} - w \sqrt{(\lambda_{12}^2 - y_{12}^2) (\lambda_{34}^2 - y_{34}^2)}\cos\phi  ) \\
    %%%%%%
    \label{eq:phi}
    \epsilon_{\mu\nu\rho\sigma}p_1^{\mu}p_2^{\nu}p_3^{\rho}p_4^{\sigma} &= 
    -\frac{\lambda M^4 w}{16}\sqrt{(\lambda_{12}^2 - y_{12}^2) (\lambda_{34}^2 - y_{34}^2)}\sin\phi,
  \end{align}
where $\bar{p}_{ij}\equiv p_i -p_j$, $\epsilon^{0123}=+1$, $y_{ij}=\lambda_{ij}\cos\theta_i^{ij}$\footnote{Their definition for $\cos\theta_{ij,k}^{ij}$ has the wrong sign, which is nevertheless of relevance for the geometrical interpretation alone. In addition, from their Eq.~(3) and Eqs.~(B1-B5), we infer that they employ $\epsilon^{0123}=-1$.} and where $z= z_{12,34}$, $w= w_{12,34}$, $\lambda= \lambda_{12,34}$.\footnote{This is, $z=(M^2-s_{12}-s_{34})M^{-2}$, $w^2=4s_{12}s_{34}M^{-4}$ and $\lambda=M^{-2}((M^2-s_{12}-s_{34})^2 -4s_{12}s_{34})^{1/2}$.} 
With this notation, the four-body phase space can be expressed as
  \begin{equation}
    d\Phi_4  =\frac{\mathcal{S}\lambda}{2^{14}\pi^{6}} \ ds_{12}ds_{34}d\phi \ \lambda_{12}d\cos{\theta}_1^{12} \ \lambda_{34}d\cos{\theta}_3^{34} 
             =\frac{\mathcal{S}\lambda}{2^{14}\pi^{6}} \ ds_{12}ds_{34} dy_{12} dy_{34} d\phi,
  \end{equation}
with $\mathcal{S}=1(1/4)$ a symmetry factor for different(identical) fermions in the final state. The integration boundaries are the following
  \begin{equation}
    4m_a^2 \leq  s_{12} \leq (M-2m_b)^2; \quad  4m_b^2 \leq  s_{34} \leq (M-\sqrt{s_{12}})^2; \quad 
    - \lambda_{ij} \leq y_{ij} \leq \lambda_{ij}; \quad  0 \leq \phi \leq 2\pi.
  \end{equation}

In addition, whenever identical leptons are present, it is useful to introduce the shorthand $ \Xi =  w \sqrt{(\lambda_{12}^2 - y_{12}^2) (\lambda_{34}^2 - y_{34}^2)}\cos\phi$ and $\eta^2=4m^2/M^2$ with $m_a=m_b\equiv m$. With these definitions, the exchange variables [noted with subscript ``$ex$'' and defined in analogy to \cref{eq:z,eq:y12,eq:y34,eq:xi,eq:phi}] read\footnote{The $\pm$ sign  in $x_{14(32)}$ is wrong in Ref.~\cite{Barker:2002ib}; that is however irrelevant since these always appear in pairs.}
  \begin{align}
    x_{14(32)}    & =  \frac{1}{4}\left(2\eta^2 +z(1-y_{12}y_{34}) \pm\lambda(y_{12}-y_{34}) +\Xi    \right), \\
    y_{14(32)}    & = \frac{1}{\lambda_{ex}}\left(\frac{\lambda}{2}(y_{12}+y_{34}) \pm (x_{12}-x_{34}) \right), \\
    \Xi_{ex}      & = z_{ex}y_{14} y_{32} - (\eta^2-x_{12}-x_{34}) - \frac{z}{2}(1+y_{12}y_{34}) +\frac{1}{2}\Xi \ ,\\
    \sin\phi_{ex} & = -\left(\frac{x_{12}x_{34}(\lambda_{12}^2 - y_{12}^2) (\lambda_{34}^2 - y_{34}^2)}{x_{14}x_{32}(\lambda_{14}^2 - y_{32}^2) (\lambda_{14}^2 - y_{32}^2)}\right)^{1/2}\sin\phi \ ,
  \end{align}
where the last two equations allow us to extract $\phi_{ex}$. This technique has been used to obtain \cref{eq:ampexc}, where the analogous of $\lambda$, $\lambda_{ex} = ((1-x_{14}-x_{32})^2 -4_{x_{14}x_{32}})^{1/2}$, has been introduced.

Finally, if one is interested in creating a MC generator, it may be useful to assign to each particle a four-momentum (in the parent particle rest frame) in terms of the phase-space variables as follows\footnote{See the axes orientation in \cref{fig:kin}.}:
  \begin{align}
    E_{1(2)} =& M\frac{1 +\delta \pm \lambda y_{12}}{4}, \qquad
    E_{3(4)} = M\frac{1 -\delta \pm  \lambda y_{34}}{4}, \\
    %%%
    \vec{p}_{1(2)} =& \mp M\sqrt{\frac{x_{12}}{4}(\lambda_{12}^2 -y_{12}^2)} \ \hat{x} \ + \
                     M\frac{\lambda \pm(1 +\delta)y_{12}}{4} \ \hat{y}, \\
    %%%
    \vec{p}_{3(4)} =& M\sqrt{\frac{x_{34}}{4}(\lambda_{34}^2 -y_{34}^2)}(\mp\cos\phi \ \hat{x} \  \pm \sin\phi \ \hat{z}) \ - \
                     M\frac{\lambda \pm(1 -\delta)y_{34}}{4} \ \hat{y} \ ,
  \end{align}
with $\delta=\delta_{12,34}$. If required, shifting among reference frames involves a Lorentz boost along the $\vec{p}_{34(12)}$ direction with parameters $\beta_{12(34)}=\lambda(1\pm\delta)^{-1}$ and $\gamma_{12(34)}=(1\pm\delta)/(2x_{12(34)})$.

%%%%%%%%%%%%%%%%%%%%%%%%%%%%%%%%%%%%%%%%%%%%%%%%%%%%%%%%%%%%%%%%%
%%%%%%%%%%%%%%%%%%%%%%%%%     CPT      %%%%%%%%%%%%%%%%%%%%%%%%%%
%%%%%%%%%%%%%%%%%%%%%%%%%%%%%%%%%%%%%%%%%%%%%%%%%%%%%%%%%%%%%%%%%
\section{\large{$\boldsymbol{C\!P}$-violating terms}}\label{sec:cpv}

The effective Lagrangian describing pseudoscalar interactions with real photons is
  \begin{equation}
  \mathcal{L}_{P\gamma\gamma} = e^2 \frac{F_{P\gamma\gamma}}{4} F^{\mu\nu}\tilde{F}_{\mu\nu}P + e^2 \frac{F_{P\gamma\gamma}^{\cancel{C\!P}}}{4}F^{\mu\nu}F_{\mu\nu}P,
  \end{equation}
where $\tilde{F}_{\mu\nu} = \frac{1}{2}\epsilon_{\mu\nu\rho\sigma}F^{\rho\sigma}$ ($\epsilon^{0123} =+1$). The first part is $C\!P$ conserving and corresponds to the LO term in chiral perturbation theory. Higher orders would modify the LO prediction for $F_{P\gamma\gamma}$ and induce a $q^2$-dependent TFF; all such effects are encoded in $F_{P\gamma\gamma}(q_1^2,q_2^2)$, and the result is valid in full generality. Concerning the $C\!P$-violating part, the most general structure features an additional gauge-invariant term~\cite{Moussallam:1994at} besides that in \cref{eq:MS}.
\footnote{
Consequently, one should modify the gauge structure in \cref{eq:MS} to 
$\left[g^{\rho\sigma}(q_{12}\cdot q_{2}) - q_{1}^{\sigma}q_{2}^{\rho}\right]F_{P\gamma\gamma}^{\cancel{C\!P}1}(q_1^2,q_2^2) + 
\left[q_1^2q_2^2 g_{\rho\sigma} + (q_1\cdot q_2)q_1^{\rho}q_2^{\sigma} - q_1^2 q_2^{\rho}q_2^{\sigma} - q_2^2 q_1^{\rho}q_1^{\sigma}\right] 
F_{P\gamma\gamma}^{\cancel{C\!P}2}(q_1^2,q_2^2)$.
}
Still, such additional structure is suppressed for quasireal photons and should play a subleading role, for which we do not include it here, but limit ourselves to correct some typos in \cite{Barker:2002ib}.\footnote{Moreover, \CP violation in double-Dalitz decays does not necessarily arise from the $P\gamma\gamma$ vertex. Another possibility is \CP violation in $P\to\bar{\ell}\ell$, which would contribute here, similar to \cref{sec:3pchpt}. We relegate therefore a more general study for later work.}
Defining the amplitudes as, $\bra{\bar{\ell}'\ell'\bar{\ell}\ell}S\ket{P}\equiv 1 +i\mathcal{M}(2\pi)^4\delta^{(4)}(P-\sum_ip_i)$~\cite{Peskin:1995ev}, the following term arises besides that in \cref{sec:LO}:
  \begin{equation}\label{eq:MS}
    i\mathcal{M}^{C\!P}_{D} =   -ie^4\frac{F_{P\gamma\gamma}^{\cancel{C\!P}}(s_{12},s_{34})}{s_{12}s_{34}} 
    \left(g^{\rho\sigma}(p_{12}\cdot p_{34}) - p_{12}^{\sigma}p_{34}^{\rho}\right) (\bar{u}_1\gamma_{\rho}v_2) (\bar{u}_3\gamma_{\sigma}v_4),
  \end{equation}
with an additional exchange amplitude if identical leptons appear (again, a relative sign would appear too). This produces the following contributions to $|\mathcal{M}|^2$ 
  \begin{align}
    \vert\mathcal{M}^{C\!P}_{D}\vert^2 +\vert\mathcal{M}^{C\!P}_{E}\vert^2   
    + 2\operatorname{Re}\left( \M{$C\!P$}{D}\M{$C\!P*$}{E} + 
    \left[\mathcal{M}^{\textrm{LO}}_{D}\mathcal{M}^{C\!P*}_{D} 
          + \mathcal{M}^{\textrm{LO}}_{D}\mathcal{M}^{C\!P*}_{E} + {D\leftrightarrow E}\right]   
    \right),  
  \end{align}
which we find to be
  \begin{multline}
    \label{eq:Sdir}
    \vert\mathcal{M}^{C\!P}_{D}\vert^2 = \frac{e^8 |F_{P\gamma\gamma}^{\cancel{C\!P}}(s_{12},s_{34})|^2}{x_{12}x_{34}} 
    \Big( z^2\Big[2 -(\lambda_{12}^2 -y_{12}^2 +\lambda_{34}^2 -y_{34}^2) + \\ (\lambda_{12}^2 -y_{12}^2)(\lambda_{34}^2 -y_{34}^2)\cos^2\phi \Big]
    -2zy_{12}y_{34}\Xi +w^2(1-y_{12}^2)(1-y_{34}^2) \Big),
  \end{multline}
  \begin{multline}\label{eq:cpvdir}
    2\operatorname{Re} \mathcal{M}^{\textrm{LO}}_{D}\mathcal{M}^{C\!P*}_{D} = 
    \frac{e^8 2\operatorname{Re} F_{P\gamma\gamma}(s_{12},s_{34})F_{P\gamma\gamma}^{\cancel{C\!P}*}(s_{12},s_{34})}{x_{12}x_{34}} \lambda\Big( 
    z(\lambda_{12}^2-y_{12}^2)(\lambda_{34}^2-y_{34}^2) \\ \times\sin\phi\cos\phi-  
    y_{12}y_{34}\sqrt{w^2(\lambda_{12}^2-y_{12}^2)(\lambda_{34}^2-y_{34}^2)}\sin\phi
    \Big),
  \end{multline}
which agrees with Ref.~\cite{Barker:2002ib} except for the $\Xi$-term sign. Moreover, we note that the overall sign from Ref.~\cite{Barker:2002ib} seems to be opposite as well given their result in Eq.~(A15), opposite to \cref{eq:phi} (see comments below). Besides, whenever identical leptons are present, the following terms appear
  \begin{multline}
    \label{eq:Sexch}
    2\operatorname{Re}\mathcal{M}^{C\!P}_{D}\mathcal{M}^{C\!P*}_{E} = 
    -\frac{e^8 \operatorname{Re}F_{P\gamma\gamma}^{\cancel{C\!P}}(s_{12},s_{34})F_{P\gamma\gamma}^{\cancel{C\!P}*}(s_{14},s_{32})}{8x_{12}x_{34}x_{14}x_{32}} 
    \Big( 8\eta^4[z-z^2-w^2y_{12}y_{34}] \\ +2\eta^2\left[ 2z^2(1+ y_{12}y_{34})(z-1) -w^2(1 -y_{12}y_{34})(2+3z[1 +y_{12}y_{34}])\right] \\
    +w^2\left[2(w^2-z^2)(1-y_{12}^2)(1-y_{34}^2) +z^2(1-y_{12}^2y_{34}^2)(1+y_{12}y_{34})  \right] \\
    +\Xi\big[ 8\eta^4z -2\eta^2(1+3y_{12}y_{34})(w^2+z^2) +z(1+y_{12}y_{34})(2w^2(-1+y_{12}y_{34}) +z^2(1+y_{12}y_{34}))\big] \\
    +\Xi^2\left[ 6\eta^2z -w^2-(2z^2+w^2)y_{12}y_{34}\right] +z\Xi^3
    \Big),
  \end{multline}
  \begin{multline}\label{eq:cpvint}
    2\operatorname{Re}\left( \mathcal{M}^{\textrm{LO}}_{D}\mathcal{M}^{C\!P*}_{E} +\mathcal{M}^{\textrm{LO}}_{E}\mathcal{M}^{C\!P*}_{D}\right) = 
    \Big[ \frac{e^8 \operatorname{Re} F_{P\gamma\gamma}(s_{12},s_{34})F_{P\gamma\gamma}^{\cancel{C\!P}*}(s_{14},s_{32})}{8x_{12}x_{34}x_{14}x_{32}} 4\lambda \\
    \times\Big( 2[x_{12}x_{34}-x_{14}x_{32}] +\eta^2[x_{12}+x_{34} +3(x_{13}+x_{24})] -x_{13}^2 -x_{24}^2 -4\eta^4  \Big) \\ 
    - (2\leftrightarrow 4) \Big] \times\sqrt{w^2(\lambda_{12}^2-y_{12}^2)(\lambda_{34}^2-y_{34}^2)}\sin\phi.
  \end{multline}
Again, the last equation differs from Ref.~\cite{Barker:2002ib}, which is only correct if both TFFs share the same $q^2$ dependency. Moreover, we note that the overall sign seems ok, but in contradiction to their result equivalent to \cref{eq:cpvdir}.

%%%%%%%%%%%%%%%%%%%%%%%%%%%%%%%%%%%%%%%%%%%%%%%%%%%%%%%%%%%%%%%%%
%%%%%%%%%%%%%%%%%%     TFF & PAs     %%%%%%%%%%%%%%%%%%%%%%%%%%%%
%%%%%%%%%%%%%%%%%%%%%%%%%%%%%%%%%%%%%%%%%%%%%%%%%%%%%%%%%%%%%%%%%
\section{\large{TFF description}}\label{sec:TFF}

There is plenty of work devoted to the study of the pseudoscalar TFF, $F_{P\gamma\gamma}(q_1^2,q_2^2)$, which is a non-perturbative object and hard to obtain from first principles. Still, given the kinematics of this process, it is mainly the low-energy regime that is required alone, yet the loop-integrals---especially the three-point ones---require a reasonable high-energy description as well. For this reason, we follow the work in Refs.~\cite{Masjuan:2012wy,Escribano:2013kba,Escribano:2015nra,Escribano:2015yup}, where the mathematical framework of Pad\'e approximants was shown to be an excellent tool to implement both regimes for the single-virtual case. This was extended to the double-virtual case in Refs.~\cite{Masjuan:2015lca,Masjuan:2015cjl,Masjuan:2017tvw} and involves the use of Canterbury approximants. 
The simplest approach\footnote{We employ factorized denominators; otherwise, the three-, four-, and five-point loop amplitudes would be hard to evaluate. If interested in the operator product expansion (OPE) behavior, one should use a model resembling that of LMD+V~\cite{Knecht:2001xc} with parameters fixed to the taylor expansion rather than masses.} reads  
  \begin{equation}\label{eq:vmd}
    F_{P\gamma\gamma}(q_1^2,q_2^2) = F_{P\gamma\gamma}\frac{M_{V_P}^2}{q_1^2-M_{V_P}^2}\frac{M_{V_P}^2}{q_2^2-M_{V_P}^2},
  \end{equation}
where $F_{P\gamma\gamma}\equiv F_{P\gamma\gamma}(0,0)$ is the normalization, that is absorbed when normalizing to $\Gamma_{P\to2\gamma}$.
It must be overemphasized that $M^2$ is not any physical vector meson mass and is related to the slope parameter. From the most updated values in Ref.~\cite{Masjuan:2017tvw} and Ref.~\cite{AlaviHarati:2001wd} for the $K_L$\footnote{We take the average result from the two parametrizations employed in Ref.~\cite{AlaviHarati:2001wd}, the Bergstr\"om-Mass\'o-Singer and the D'Ambrosio-Isidori-Portol\'es models, each of them leading to $0.59(2)$~GeV and $0.62(2)$~GeV for $M_{K_L}$, respectively.} we find 
  \begin{align}
    M_{V_{\pi^0}} = 0.754(23)\textrm{GeV} ;\ \ M_{V_{\eta}} = 0.724(5)\textrm{GeV} ;\ \   M_{V_{\eta'}} = 0.837(10)\textrm{GeV} ;\ \   M_{V_{K_L}} 0.61(2)\textrm{GeV}.
  \end{align}
When evaluating some loop amplitudes, expressions containing $F_{P\gamma\gamma}(q_1^2,q_2^2)(q_1^2+i\epsilon)^{-1} (q_2^2+i\epsilon)^{-1}$ appear. In order to evaluate the integrals, it is useful to use partial fraction decomposition  that, for \cref{eq:vmd}, reads
  \begin{equation}
    \frac{F_{P\gamma\gamma}(q_1^2,q_2^2)}{q_1^2 q_2^2} = \frac{F_{P\gamma\gamma}}{q_1^2 q_2^2} 
        -\frac{F_{P\gamma\gamma}}{(q_1^2 - M_{V_{P}}^2) q_2^2}  -\frac{F_{P\gamma\gamma}}{q_1^2 (q_2^2 - M_{V_{P}}^2)}  
        +\frac{F_{P\gamma\gamma}}{(q_1^2 - M_{V_{P}}^2) (q_2^2 - M_{V_{P}}^2)}.
  \end{equation}
As a consequence, the loop integrals can be evaluated for arbitrary photon masses $M_{V_{P}}$ and a constant TFF; the full result is obtained by adding the four terms above, which is implicit in the main text. If employing a more elaborated TFF, the procedure is analogous and would produce additional terms.

%%%%%%%%%%%%%%%%%%%%%%%%%%%%%%%%%%%%%%%%%%%%%%%%%%%%%%%%%%%%%%%%%
%%%%%%%%%%%%%%%%     BREMSSTRAHLUNG     %%%%%%%%%%%%%%%%%%%%%%%%%
%%%%%%%%%%%%%%%%%%%%%%%%%%%%%%%%%%%%%%%%%%%%%%%%%%%%%%%%%%%%%%%%%
\section{\large{Bremsstrahlung integral}}\label{sec:bsint}

The solution to \cref{eq:bsint} has been given in Ref.~\cite{tHooft:1978jhc}. The general result reads 
  \begin{align}
  J(p_i,p_j) = &\frac{1}{(2\pi)^2}\frac{1}{p_{ij}^2\lambda_{i,j}} \Bigg[  
  \ln\left(\frac{z_{i,j}+\lambda_{i,j}}{z_{i,j}-\lambda_{i,j}}\right)\ln\left(\frac{2E_c}{m_{\gamma}}\right)  \nonumber\\ 
   & +\frac{1}{4}\ln^2\left(\frac{u^0-\boldsymbol{u}}{u^0+\boldsymbol{u}}\right)  
   + \operatorname{Li_2}\left(1 - \frac{u^0-\boldsymbol{u}}{v} \right) 
   + \operatorname{Li_2}\left(1 - \frac{u^0+\boldsymbol{u}}{v} \right) \Bigg\vert^{u=\alpha p_i}_{u=p_j} \label{eq:BSInt}
  \Bigg]. 
\end{align}
where $\alpha = (2p_i\cdot p_j + p_{ij}^2\lambda_{ij})/(2m_i^2)$ and $v=(\alpha^2p_i^2-p_j^2)/(2(\alpha p_i^0 -p_j^0))$. In order to associate the cutoff energy, $E_c$, with the $4\ell$ momenta, the parent particle frame should be adopted to evaluate the expression above. Note that in the soft photon approximation this coincides with the $4\ell$ rest frame. We find that, using the notation in \cref{sec:kin},
  \begin{align}
    p_i^0 ={}& M\frac{1+\delta_{i,jkl}}{2}; \ \ p_j^0 = M\frac{1+\delta_{j,ikl}}{2};  \ \ 
        \boldsymbol{p}_i =  M\frac{\lambda_{i,jkl}}{2};  \ \ \boldsymbol{p}_j = M\frac{\lambda_{j,ikl}}{2}, \\
    \alpha ={}& \frac{z_{ij}+\lambda_{ij}}{1-z_{ij}+\delta_{i,j}}  \equiv \sigma_{ij}; \quad 
      v = \frac{\sigma_{ij} p_{ij}^2\lambda_{i,j}}{M(\sigma_{ij}(1+\delta_{i,jkl}) - (1+\delta_{j,ikl}))}
      \equiv M\frac{\sigma_{ij} x_{ij}\lambda_{i,j}}{\Upsilon_{ij}}, \\
    \Omega_{i}^{\pm} \equiv & \frac{1}{M}(p_{i}^0 \pm \boldsymbol{p}_{i}) =  
      \frac{1}{2}(1+\delta_{i,jkl} \pm\lambda_{i,jkl}); \quad 
    \Omega_{j}^{\pm} \equiv  \frac{1}{M\alpha}(p_{j}^0 \pm \boldsymbol{p}_{j}) =  
      \frac{1}{2\sigma_{ij}}(1+\delta_{j,ikl} \pm\lambda_{j,ikl}),
  \end{align}
in analogy with Ref.~\cite{Barker:2002ib}. Furthermore, we give below the particular value for the new variables that are required in terms of phase space ones,
\begin{align}
x_{14(23)}z_{14(23)} = \frac{ z(1-y_{12}y_{34}) \pm\lambda(y_{12} - y_{34}) +\Xi }{4}, %\quad
    x_{14(23)}\lambda_{14(23)} = \sqrt{x_{14(23)}^2 z_{14(23)}^2 - \frac{4m_a^2m_b^2}{M^4}}\\
x_{13(24)}z_{13(24)} =\frac{ z(1+y_{12}y_{34}) \pm\lambda(y_{12} + y_{34}) -\Xi}{4},  %\quad
    x_{13(24)}\lambda_{13(24)} = \sqrt{x_{13(24)}^2z_{13(24)}^2 - \frac{4m_a^2m_b^2}{M^4}},\\
\delta_{^{1,234}_{2,134}} = -\frac{ 1 -x_{12} +x_{34} \mp \lambda y_{12} }{2}, \
\lambda_{^{1,234}_{2,134}} =  \frac{\sqrt{\lambda^2(1 +y_{12}^2) + 4x_{12}\lambda_{12}^2 \pm 2\lambda y_{12}(1 +x_{12} -x_{34})}}{2}, 
\end{align}
with the remaining $\{\delta,\lambda\}_{^{3,124}_{4,123}}$ combinations obtained by replacing $(12)\leftrightarrow(34)$. Note, particularly, that $x_{ij}z_{ij}$ and $x_{ij}\lambda_{ij}$ can be employed instead of $z_{ij}, \lambda_{ij}$ which are more involved.

%%%%%%%%%%%%%%%%%%%%%%%%%%%%%%%%%%%%%%%%%%%%%%%%%%%%%%%%%%%%%%%%%
%%%%%%%%%%%%%%%%%%       3-POINT       %%%%%%%%%%%%%%%%%%%%%%%%%%
%%%%%%%%%%%%%%%%%%%%%%%%%%%%%%%%%%%%%%%%%%%%%%%%%%%%%%%%%%%%%%%%%
\section{\large{Three-point amplitudes in $\boldsymbol{\chi}$PT}}\label{sec:3pchpt}

\begin{figure}
    \includegraphics[width=\textwidth]{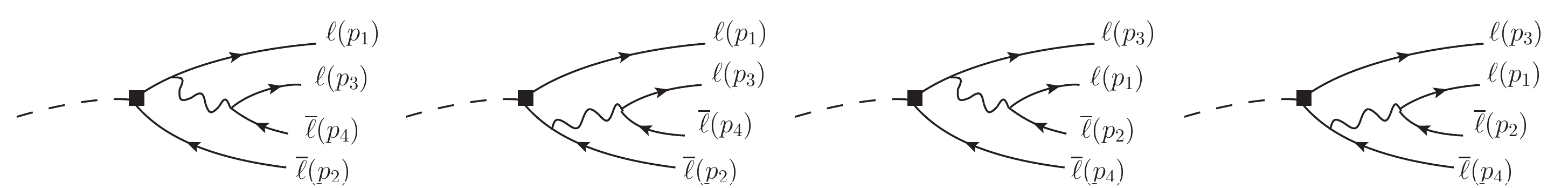}
    \caption{Counterterm diagrams from $\chi$PT. Again, additional diagrams arise if identical leptons appear.\label{fig:chpt}}
  \end{figure}

For a constant TFF---which would correspond to the LO in the chiral expansion---the three-point integrals are divergent. Particularly, for $\MM{3P}{1D}$, we find that\footnote{In particular, using dimensional regularization in $d=4-\epsilon$ dimensions, the divergence for the given integral reads $-(3/2)\Delta_{\epsilon}(\mu)$, with $\Delta_{\epsilon}(\mu) = 2\epsilon^{-1} -\gamma_E +\ln(4\pi\mu^2)$, with $\mu$ the renormalization scale. Note that DR entails an additional $1/4$ term absent in other regularization schemes.}
  \begin{multline}\label{eq:div}
    \operatorname{Div}  i\MM{3P}{1D} = \frac{-e^2\alpha^2F_{P\gamma\gamma}}{p_{34}^2(p_{134}^2-m_a^2)}
    (\overline{u}_1\gamma^{\lambda}(\slashed{p}_{134}+m_a)\slashed{P}\gamma^5 v_2) (\overline{u}_3 \gamma_{\lambda} v_4) \\ 
    \times \operatorname{Div} \frac{2i}{\pi^2P^2} \int d^4k \frac{P^2k^2(1-d^{-1})}{k^2(k+P)^2((k+p_2)^2-m_a^2)},
  \end{multline}
with obvious results for the additional amplitudes. The loop-integral divergence must cancel when including the appropriate counterterm. This is the same as that appearing in $P\to\bar{\ell}\ell$ decays, introduced in Ref.~\cite{Savage:1992ac}, and which in this process manifests as
  \begin{equation}
     \mathcal{L}_{\chi\textrm{PT}} \supset \chi(\mu)\alpha^2F_{P\gamma\gamma}^{\textrm{LO}}(\overline{\ell}\gamma^{\mu}\gamma^5\ell)\partial_{\mu}P, \qquad \chi(\mu) \equiv -(\chi_1(\mu) + \chi_2(\mu))/4,
  \end{equation}
with $\mu$ the renormalization scale.\footnote{It is a common practice to use $\chi(0.77)$; for an arbitrary scale $\mu$, $\chi(\mu) = \chi(0.77) + 3\ln(\mu/0.77)$ with $\mu$ in GeV.} This produces the following amplitudes appearing in \cref{fig:chpt}
  \begin{align}
%%%%%%%%%%%%%%%%%%1d
i\mathcal{M}_{\textrm{1D}}^{\chi} ={}& 
\frac{-e^2\alpha^2F_{P\gamma\gamma}}{p_{34}^2(p_{134}^2-m_a^2)} \chi(\mu) (\overline{u}_1\gamma^{\lambda}(\slashed{p}_{134}+m_a)\slashed{P}\gamma^5 v_2) 
(\overline{u}_3 \gamma_{\lambda} v_4), \\
%%%%%%%%%%%%%%%%%%2d
i\mathcal{M}_{\textrm{2D}}^{\chi} ={}& 
\frac{-e^2\alpha^2F_{P\gamma\gamma}}{p_{34}^2(p_{234}^2-m_a^2)} \chi(\mu) (\overline{u}_1\slashed{P}\gamma^5(-\slashed{p}_{234}+m_a)\gamma^{\lambda} v_2) 
(\overline{u}_3 \gamma_{\lambda} v_4), \\
%%%%%%%%%%%%%%%%%%3d
i\mathcal{M}_{\textrm{3D}}^{\chi} ={}& 
\frac{-e^2\alpha^2F_{P\gamma\gamma}}{p_{12}^2(p_{123}^2-m_b^2)} \chi(\mu) (\overline{u}_3\gamma^{\lambda}(\slashed{p}_{123}+m_b)\slashed{P}\gamma^5 v_2) 
(\overline{u}_1 \gamma_{\lambda} v_2), \\
%%%%%%%%%%%%%%%%%%4d
i\mathcal{M}_{\textrm{4D}}^{\chi} ={}& 
\frac{-e^2\alpha^2F_{P\gamma\gamma}}{p_{12}^2(p_{124}^2-m_b^2)} \chi(\mu) (\overline{u}_3\slashed{P}\gamma^5(-\slashed{p}_{124}+m_b)\gamma^{\lambda} v_2) 
(\overline{u}_1 \gamma_{\lambda} v_2), 
\end{align}
and corresponding exchange amplitudes whenever identical leptons are present. In the light of the equations above and \cref{eq:div}, it is clear that divergences cancel exactly in the same manner as in $P\to\bar{\ell}\ell$ decays~\cite{Masjuan:2015cjl} (see Eq.~(2.2) and Eq.~(6.1) therein) as it should be. Concerning the NLO correction, it shifts $\mathcal{I}_1\to\mathcal{I}_1 -\chi(\mu)/2$ in \cref{eq:nlo3p,eq:3pint1,eq:3pint2}. At this order, the same counterterm applies to $\pi^0, \eta$, $\eta'$ and, essentially, to $K_L$ as well (see Ref.~\cite{GomezDumm:1998gw}). This may not be appropriate however---see discussions in Ref.~\cite{Masjuan:2015cjl}---as it would produce different counterterms for each pseudoscalar and lepton species.\footnote{In Ref.~\cite{Masjuan:2015cjl} it was shown that different pseudoscalars ($P=\pi^0,\eta$), TFFs (Fact vs OPE there), and leptonic channels ($\ell=e,\mu$) lead $\chi\in(2.53\div 6.46)$. For the $K_L$ it would give $\chi\in(6.68\div 7.25)$ and $\chi\in(3.74\div 4.28)$ for $\ell=e,\mu$ and (Fact$\div$OPE).}
In order to show the accuracy of the chiral expansion, we give Dir$^{\textrm{NLO}}$ numerically in terms of $\chi$. For such purpose, it is convenient to express it as
  \begin{equation}\label{eq:3pdec}
    \Gamma^{\textrm{NLO}}_{\textrm{3P;D}}/\Gamma^{\textrm{LO}}  \equiv \delta_{\textrm{UV}}(0.77) +  \sum_{\ell} \delta_{\ell}\chi(0.77),
  \end{equation}
where summation is meant for $e^+e^-\mu^+\mu^-$ cases alone, and coefficients, $\delta_{\textrm{UV},\ell}$, given in \cref{tab:3pchpt}.
From the results therein, it is clear that counterterm effects are irrelevant for the purely electronic channels. For channels including muons, there is however a delicate cancellation among the loop and counterterms, which makes this contribution quite sensitive to $\chi(0.77)$, in contrast to the calculation including the TFF. To find a better agreement with the latter, we find it better to use the $\chi$ associated to the same pseudoscalar and lepton from Ref.~\cite{Masjuan:2015cjl}. Moreover, we found it better to adopt our results in \cite{Masjuan:2015cjl} corresponding to a factorized TFF. Indeed, we employed a more elaborate result for the TFF concerning three-point corrections and found it irrelevant to include the OPE or not, in contrast to $P\to\bar{\ell}\ell$ decays.
  \begin{table}[tbp]\centering\footnotesize
  \begin{tabular}{cccccccc}\toprule
           & $\pi^0\to 4e$ & $K_L\to 4e$ & $K_L\to 2e2\mu$ & $K_L\to 4\mu$ & $\eta\to 4e$ & $\eta\to 2e2\mu$ & $\eta\to 4\mu$  \\ \midrule
  $\delta_{\textrm{UV}}(0.77)$& $-0.0007(0)$& $-0.0010(1)$& $\phantom{-}0.00020(1)$& $\phantom{-}0.0063(3)$& $-0.0010(1)$& $\phantom{-}0.00020(1)$& $\phantom{-}0.0321(13)$ \\ 
  $\delta_e$                  & $-0.0000(0)$& $-0.0000(0)$& $-0.00000(0)$          & $-$                   & $-0.0000(0)$& $-0.00000(0)$          & $-$ \\ 
  $\delta_{\mu}$              & $-$         & $-$         & $-0.00005(1)$          & $-0.0018(1)$          & $-$          & $-0.00006(1)$          & $-0.0101(3)\phantom{1}$ \\ \bottomrule
  \end{tabular}
  \caption{Numerical values for the terms in \cref{eq:3pdec} in same units as \cref{tab:LO}.\label{tab:3pchpt}}
  \end{table}

%%%%%%%%%%%%%%%%%%%%%%%%%%%%%%%%%%%%%%%%%%%%%%%%%%%%%%%%%%%%%%%%%
%%%%%%%%%%%%%%%%%%       TABLES        %%%%%%%%%%%%%%%%%%%%%%%%%%
%%%%%%%%%%%%%%%%%%%%%%%%%%%%%%%%%%%%%%%%%%%%%%%%%%%%%%%%%%%%%%%%%
\clearpage\section{\large{Numerical NLO corrections}}\label{sec:NLOtables}

  \begin{table}[h!]\centering\tiny
  \begin{tabular}{cccccccc} \toprule 
  \multicolumn{8}{c}{$\pi^0\to e^+e^-e^+e^-$} \\ \midrule
  D+E&  $\phantom{-}0.0389(2)$ & $-0.0032(2)$ & $-0.6355(6)$ & $-0.0007(0)$ &  $-0.0016(1)$ & $0$ & $-0.6021(7)$\\
  Int&  $-0.0008(1)$& $\phantom{-}0.0000(0)$& $\phantom{-}0.0123(1)$& $-0.0004(0)$& $\phantom{-}0.0005(0)$& $0.0009(1)$& $\phantom{-}0.0125(2)$\\ 
  Total&$\phantom{-}0.0381(2)$& $-0.0032(2)$& $-0.6232(6)$& $-0.0011(0)$& $-0.0011(1)$& $0.0009(1)$& $-0.5896(7)$\\  \toprule
  %%%%
  \multicolumn{8}{c}{$K_L^0\to e^+e^-e^+e^-$} \\ \midrule
  D+E &  $\phantom{-}0.0942(1)$ & $-0.0043(0)$ & $-1.5142(4)$  & $-0.0010(0)$&  $-0.0022(2)$ & $0$ & $-1.4275(4)$\\
  Int&$-0.0010(1)$&$\phantom{-}0.0000(0)$&$\phantom{-}0.0166(2)$&$-0.0007(0)$&$\phantom{-}0.0007(0)$&$\phantom{-}0.0010(2)$&$\phantom{-}0.0166(3)$\\ 
  Total& $\phantom{-}0.0932(2)$& $-0.0043(0)$& $-1.4979(4)$& $-0.0017(0)$& $-0.0015(2)$& $\phantom{-}0.0010(2)$& $-1.4109(5)$\\  \toprule
  %%%%
  \multicolumn{8}{c}{$K_L^0\to e^+e^-\mu^+\mu^-$} \\ \midrule
  Dir    & $0.0621(1)$ & $-0.0065(0)$ & $-0.2748(3)$ & $-0.0025(1)$ & $-0.0063(4)$ & $0$ & $-0.2280(5)$ \\
  \toprule
  %%%%
  \multicolumn{8}{c}{$K_L^0\to \mu^+\mu^-\mu^+\mu^-$} \\ \midrule
  D+E & $\phantom{-}0.0258(0)$& $-0.0043(0)$& $\phantom{-}0.0536(0)$& $-0.0013(0)$& $-0.0024(0)$& $0$ & $\phantom{-}0.0714(1)$\\
  Int & $-0.0014(0)$& $\phantom{-}0.0002(0)$ & $-0.0023(0)$& $\phantom{-}0.0005(0)$& $-0.0001(0)$ & $-0.0057(1)$  & $-0.0087(1)$\\ 
  Total & $\phantom{-}0.0244(0)$& $-0.0041(0)$& $\phantom{-}0.0513(0)$& $-0.0008(1)$& $-0.025(0)$& $-0.0057(1)$ & $\phantom{-}0.0628(1)$\\  \toprule
  %%%%
  \multicolumn{8}{c}{$\eta\to e^+e^-e^+e^-$} \\ \midrule
  D+E& $\phantom{-}0.0996(1)$& $-0.0044(0)$& $-1.6000(5)$& $-0.0010(1)$ & $-0.0023(2)$ & $0$  & $-1.5081(5)$\\
  Int& $ -0.0010(1)$& $\phantom{-}0.0000(0)$& $\phantom{-}0.0169(2)$ & $-0.0007(1)$ & $\phantom{-}0.0007(0)$& $0.0024(2)$& $\phantom{-}0.0183(3)$\\ 
  Total& $\phantom{-}0.0986(1)$ & $-0.0044(0)$ & $-1.5831(5)$ & $-0.0017(1)$ & $-0.0016(1)$ & $0.0024(2)$& $-1.4898(6)$\\ \toprule 
  %%%%
  \multicolumn{8}{c}{$\eta\to e^+e^-\mu^+\mu^-$}\\ \midrule
  D     & $0.0890(1)$ & $ -0.0088(0)$ & $-0.4141(4)$& $-0.0026(1)$ & $-0.0087(1)$ & $0$  & $-0.3452(4)$\\ \toprule
  %%%%
  \multicolumn{8}{c}{$\eta\to \mu^+\mu^-\mu^+\mu^-$} \\ \midrule
  D+E & $\phantom{-}0.1790(2)$& $-0.0275(0)$& $\phantom{-}0.2324(2)$& $-0.0065(2)$& $-0.0147(5)$& $0$& $\phantom{-}0.3627(6)$\\
  Int & $ -0.0137(1)$& $\phantom{-}0.0021(1)$& $-0.0124(1)$& $\phantom{-}0.0009(0)$& $\phantom{-}0.0000(0)$ & $-0.0347(3)$  & $-0.0578(4)$\\ 
  Total& $\phantom{-}0.1653(2)$& $-0.0254(1)$& $\phantom{-}0.2200(2)$& $-0.0056(2)$& $-0.0147(5)$& $-0.0347(3)$& $\phantom{-}0.3049(7)$\\  \toprule
  %%%%
  & VP & $F_2$ & $F_1$ & 3P & 4P  & 5P & NLO \\ \bottomrule
  \end{tabular}
  \caption{Results for different NLO contributions with a constant TFF. Note that $F_1$ and 5P include BS contributions. The units are chosen analogous to \cref{tab:LO}.\label{tab:NLOdetail}}
  \end{table}
%%%%%%%%%%%%%%%%%%%%%%%%%%%%%%%%%%%%%
\begin{table}[h!]\centering\tiny
\begin{tabular}{cccccccc} \toprule 
\multicolumn{8}{c}{$\pi^0\to e^+e^-e^+e^-$} \\ \midrule
 D+E    & $\phantom{-}0.0392(2)$ & $-0.0032(2)$ & $-0.6391(6)$ & $-0.0007(0)$& $-0.0017(1)$ & $0$ & $-0.6055(7)$ \\
 Int & $-0.0008(1)$ & $\phantom{-}0.0000(0)$ & $\phantom{-}0.0126(1)$ & $-0.0004(0)$& $\phantom{-}0.0005(0)$ & $0.0009(1)$ & $\phantom{-}0.0128(2)$  \\ 
 Total & $\phantom{-}0.0384(2)$ & $-0.0032(2)$ & $-0.6265(6)$ & $-0.0011(0)$ & $-0.0012(1)$ & $0.0009(1)$ & $-0.5927(7)$ \\  \toprule
%%%%
\multicolumn{8}{c}{$K_L^0\to e^+e^-e^+e^-$} \\ \midrule
 D+E & $\phantom{-}0.1047(1)$ & $-0.0045(0)$ & $-1.6890(5)$  & $-0.0016(1)$ & $-0.0048(5)$ & $0$& $-1.5952(7)$  \\
 Int & $-0.0016(1)$  & $\phantom{-}0.0000(0)$ & $\phantom{-}0.0265(3)$ & $-0.0012(1)$ & $\phantom{-}0.0013(1)$ & $\phantom{-}0.0017(3)$ & $\phantom{-}0.0267(4)$  \\ 
 Total  & $\phantom{-}0.1031(1)$ & $-0.0045(0)$ & $-1.6625(6)$ & $-0.0028(1)$ & $-0.0035(5)$ & $\phantom{-}0.0017(3)$ & $-1.5685(9)$ \\  \toprule
%%%%
\multicolumn{8}{c}{$K_L^0\to e^+e^-\mu^+\mu^-$} \\ \midrule
D     & $0.1067(1)$ & $-0.0107(1)$ & $-0.4763(5)$ & $-0.0067(2)$ & $-0.0209(2)$ & $0$ & $-0.4079(8)$ \\
\toprule
%%%%
\multicolumn{8}{c}{$K_L^0\to \mu^+\mu^-\mu^+\mu^-$} \\ \midrule
D+E & $\phantom{-}0.0481(0)$& $-0.0080(0)$& $\phantom{-}0.0985(1)$& $-0.0027(1)$& $-0.0070(2)$& $0$& $\phantom{-}0.1289(2)$ \\
Int & $-0.0026(2)$& $\phantom{-}0.0004(0)$ & $-0.0044(0)$& $-0.0013(0)$& $-0.0007(0)$ & $-0.0142(1)$& $-0.0228(2)$ \\ 
Total & $\phantom{-}0.0455(2)$& $-0.0076(0)$& $\phantom{-}0.0941(1)$& $-0.0040(1)$& $-0.0077(2)$& $-0.0142(1)$& $\phantom{-}0.1061(3)$ \\  \toprule
%%%%
\multicolumn{8}{c}{$\eta\to e^+e^-e^+e^-$} \\ \midrule
 D+E & $\phantom{-}0.1086(1)$ & $-0.0045(0)$ & $-1.7490(5)$  & $-0.0016(1)$ & $-0.0044(4)$ & $0$& $-1.6509(6)$  \\
 Int & $-0.0015(1)$  & $\phantom{-}0.0000(0)$ & $\phantom{-}0.0251(2)$ & $-0.0011(1)$ & $\phantom{-}0.0012(1)$ & $\phantom{-}0.0015(2)$ & $\phantom{-}0.0016(6)$  \\ 
 Total  & $\phantom{-}0.1070(1)$ & $-0.0045(0)$ & $-1.7239(5)$ & $-0.0027(1)$ & $-0.0032(4)$ & $\phantom{-}0.0015(2)$ & $-1.6509(6)$ \\  \toprule
%%%%
\multicolumn{8}{c}{$\eta\to e^+e^-\mu^+\mu^-$} \\ \midrule
D     & $0.1337(1)$ & $ -0.0127(1)$ & $-0.6267(6)$& $-0.0057(1)$ & $-0.0224(2)$ & $0$ & $-0.5338(7)$ \\
\toprule
%%%%
\multicolumn{8}{c}{$\eta\to \mu^+\mu^-\mu^+\mu^-$} \\ \midrule
D+E & $\phantom{-}0.2914(3)$& $-0.0446(0)$& $\phantom{-}0.3679(4)$& $-0.0111(3)$& $-0.0361(11)$& $0$& $\phantom{-}0.5675(12)$ \\
Int & $ -0.0229(2)$& $\phantom{-}0.0035(1)$& $-0.0207(2)$& $-0.0056(2)$& $-0.0018(1)$ & $-0.0718(6)$& $-0.1193(7)$ \\ 
Total & $\phantom{-}0.2685(4)$& $-0.0411(1)$& $\phantom{-}0.3472(4)$& $-0.0167(4)$& $-0.0379(11)$& $-0.0718(6)$& $\phantom{-}0.4482(15)$ \\ \toprule
  & VP & $F_2$ & $F_1$ & 3P & 4P  & 5P & NLO \\ \bottomrule
\end{tabular}
\caption{Analogous results to \cref{tab:NLOdetail} for the $q^2$-dependent TFFs introduced in \cref{sec:TFF}.\label{tab:NLOdetailFF}}
\end{table}

%BIBLIOGRAPHY
\bibliographystyle{apsrev4-1}
\bibliography{references}

\end{document}